\documentclass[sigconf]{acmart}

\usepackage{url}
\usepackage{algorithm}
\usepackage{algpseudocode}

\usepackage{subcaption}

\usepackage{color} 
\usepackage{colortbl} 
\usepackage{float}

\usepackage{multirow}
\usepackage{array}

\AtBeginDocument{%
  }

\acmConference[WORKS 2023]{18th Workshop on Workflows in Support of Large-Scale Science}{November 12 2023}{Denver, CO}
\acmPrice{15.00}
\acmISBN{978-1-4503-XXXX-X/18/06}

\newcommand{\dfp}{\texttt{dispel4py}}

\newcommand{\ptime}{\textit{process time}}
\newcommand{\rtime}{\textit{runtime}}

\newcommand{\dyn}{\textit{dynamic scheduling}}
\newcommand{\naive}{\textit{naive assignment}}
\newcommand{\stage}{\textit{staging}}
\newcommand{\hybrid}{\textit{hybrid}}
\newcommand{\auto}{\textit{auto-scaling}}

\newcommand{\multiprocessing}{\textit{Multiprocessing}}
\newcommand{\redis}{\textit{Redis}}
\newcommand{\mpi}{\textit{MPI}}
\newcommand{\simple}{\textit{Simple}}
\newcommand{\storm}{\textit{Storm}}

\newcommand{\multi}{\textit{multi}}

\newcommand{\dynmulti}{\textit{dyn\_multi}}
\newcommand{\dynredis}{\textit{dyn\_redis}}
\newcommand{\automulti}{\textit{dyn\_auto\_multi}}
\newcommand{\autoredis}{\textit{dyn\_auto\_redis}}
\newcommand{\hyredis}{\textit{hybrid\_redis}}

\newcommand{\wfint}{Internal Extinction of Galaxies}
\newcommand{\wfcorr}{Seismic Cross-Correlation}
\newcommand{\wfsent}{Sentiment Analyses for News Articles}

\newcommand{\server}{\textit{server}}
\newcommand{\cloud}{\textit{cloud}}
\newcommand{\hpc}{\textit{HPC}}

\begin{document}

\title{Optimization towards Efficiency and Stateful of dispel4py}

\author{%
Liang Liang\textsuperscript{*}, Heting Zhang\textsuperscript{†}, Guang Yang\textsuperscript{*}, Thomas Heinis\textsuperscript{*}, Rosa Filgueira\textsuperscript{†}\\
* Imperial College London \\† University of St Andrews
}

\renewcommand{\shortauthors}{}
\begin{abstract}

Scientific workflows bridge scientific challenges with computational resources. While \dfp{}, a stream-based workflow system, offers mappings to parallel enactment engines like MPI or Multiprocessing, its optimization primarily focuses on dynamic process-to-task allocation for improved performance. An efficiency gap persists, particularly with the growing emphasis on conserving computing resources. Moreover, the existing dynamic optimization lacks support for stateful applications and grouping operations.

To address these issues, our work introduces a novel hybrid approach for handling stateful operations and groupings within workflows, leveraging a new Redis mapping. We also propose an auto-scaling mechanism integrated into \dfp{}'s dynamic optimization. Our experiments showcase the effectiveness of auto-scaling optimization, achieving efficiency while upholding performance. In the best case, auto-scaling reduces \dfp{}'s runtime to 87\% compared to the baseline, using only 76\% of process resources. Importantly, our optimized stateful \dfp{} demonstrates a remarkable speedup, utilizing just 32\% of the runtime compared to the contender.



\end{abstract}






\maketitle


\section{Introduction}

Powerful computing infrastructures and platforms have evolved to handle the increase in large-scale data and highly computing-intensive applications \cite{liu2015survey}. However, navigating these computing resources to tackle data-intensive scientific problems can be overwhelming for scientists from various disciplines. The challenge lies not just in choosing the proper infrastructure but also in understanding and managing it. Therefore, workflow communities are promoted as a means to design and propose various workflow systems, effectively concealing intricate technical details, in order to bridge the divide between scientific challenges and computing technologies ~\cite{atkinson2017scientific}. Workflow systems automatically handle low-level computing processing, allowing scientists to focus on their research.

\begin{figure}
    \centering
    \includegraphics[width=0.8\linewidth]{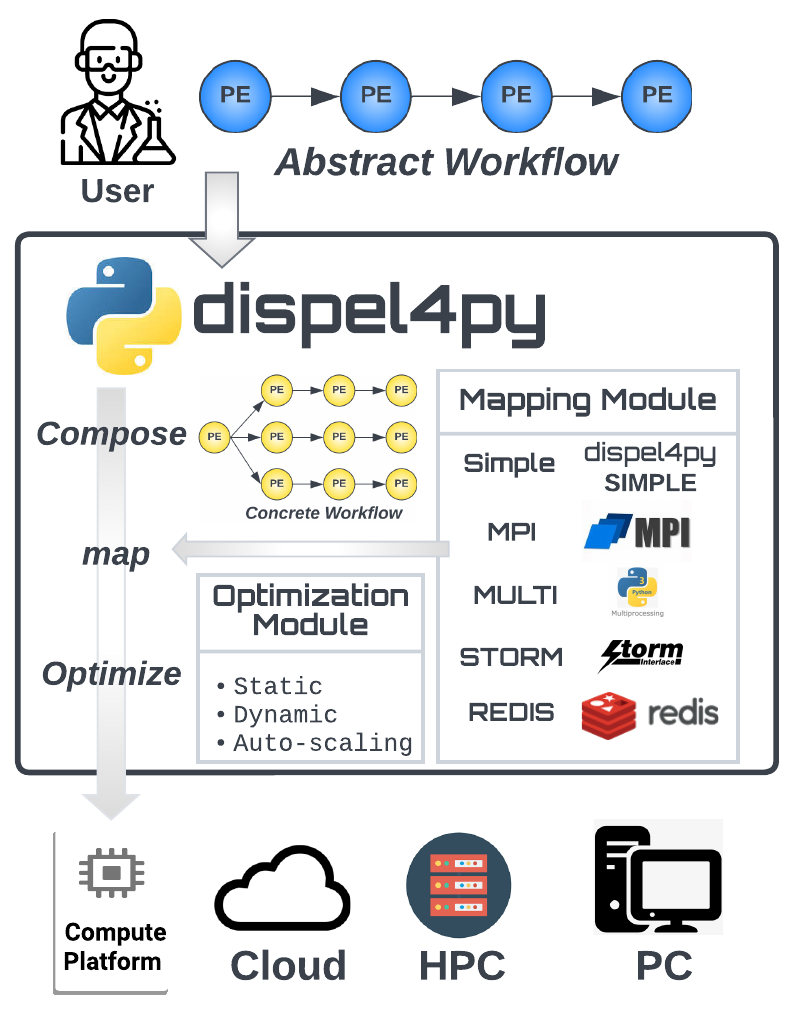}
    \caption{\dfp{} overview. From abstract workflow design to automatic concrete workflow generation and execution.}
    \label{fig:dfp_framework}
    \vspace{-15pt}
\end{figure}

When building workflows within \dfp{}, users engage in the design, composition, and interconnection of various processing elements (PEs), which serve as the fundamental computational building blocks of the system~\cite{filguiera2017dispel4py}. These PEs are linked together by users to form graphs, also known as abstract workflows. 

Subsequently, \dfp{} smoothly converts the composed abstract workflows into concrete implementations using the chosen mapping (or enactment engine). These concrete implementations are then executed on the specified computing platform, as depicted in Figure~\ref{fig:dfp_framework}. This process is underscored by the strategic decoupling of abstract workflows from underlying communication mechanisms, empowering adaptability across diverse computing resources and resonating with \dfp{}'s intuitive processing methodology that aligns harmoniously with the principles of streaming computing, capturing significant interest from researchers for its adeptness in managing dynamic and time-sensitive data \cite{filguiera2017dispel4py, filgueira2015dispel4py}.


While \dfp{} has substantial merits, performance limitations are evident. A notable concern is its basic and static workload allocation, where processing elements (PEs) are pre-assigned to available \textit{computing resources} in a static deployment of workflows. To address these challenges, previous efforts \cite{liang2022scalable, liang2020adaptive} introduced static optimizations (\naive{} and \stage{}) and a dynamic optimization (\dyn{}). \naive{} and \stage{} strategies optimize resource allocation for static deployment, while \dyn{} enhances adaptability by dynamically distributing resources among PEs, eliminating operational halts, collectively enhancing scalability and adaptability.

In the current context of increasing significance towards efficiency, encompassing performance and cost-effectiveness, consideration of monetary and energy costs is crucial. Aligning with sustainability and green computing principles, efficient systems lead to significant energy savings and reduced carbon footprint \cite{saha2014green, bharany2022systematic}. Auto-scaling emerges as a pertinent solution, dynamically adjusting resources based on real-time demands, ensuring avoidance of underutilization or over-provisioning \cite{verma2021auto}.  Driven by these principles, this work introduces auto-scaling techniques to elevate \dfp{}'s capabilities, dynamically adjusting resources in response to real-time requirements. This advancement empowers \dfp{} to enhance its efficiency, conserve energy, and bolster its adaptability to contemporary demands.

Moreover, a rising trend in stateful applications \cite{de2021distributed} presents a challenge. While \dfp{}'s static deployment accommodate such applications, compatibility with \dyn{} becomes complex when processes randomly handle tasks, potentially causing inconsistent task states. To address this, we propose the \hybrid{} method, enabling \dyn{} mappings to proficiently manage both stateless and stateful applications within \dfp{}. In essence, our contributions seamlessly integrate the \auto{} and \hybrid{} techniques, enhancing \dfp{}'s versatility, and leveraging three real-world workflows for performance evaluation.

The rest of the paper is structured as follows. Section~\ref{sec:background} provides the necessary background context. Section~\ref{sec:proposed} delves into the introduced mappings and optimizations. The use cases are presented in Section~\ref{sec:usecase}, while Section~\ref{sec:exp} highlights the  potential and effectiveness of our introduced techniques. Lastly, Section~\ref{sec:concl} concludes with final thoughts.

\section{Background}
\label{sec:background}

\subsection{\dfp{}}
\label{sec:d4py_background}


\dfp{} is a stream-based data-intensive workflow written in Python \cite{filguiera2017dispel4py}. Key concepts in \dfp{} include:

\begin{itemize}

\item \textit{Processing elements (PEs)} represent computational entities responsible for task processing or data transformation within the workflow graph. In the context of \dfp{}, various types of PEs are available. Unlike file-based interactions common in task-based workflow systems, data is seamlessly exchanged between connected PEs in a streaming fashion.

\item \textit{Instance} denotes an executable copy of a PE. A single PE can have multiple instances depending on the configuration and the number of processes.

\item \textit{Connection} transmits data from one PE instance's output port to one or more input ports of another PE instance. 

\item \textit{Mapping} is the process of `translating' workflows onto execution systems. This encompasses \simple{} mapping for sequential workflow execution, \mpi{}, alongside parallel alternatives like \multiprocessing{}\footnote{\url{https://docs.python.org/3/library/multiprocessing.html}}, \mpi{}~\cite{openmpi} and \storm{}~\cite{storm}.  Those mappings eliminate the need for manual workflow modifications.

\item \textit{Abstract workflow} consists of several PEs in the form of a directed acyclic graph (DAG), which is designed and defined by the user for solving specific problems.

\item \textit{Concrete Workflow}, also referred to as the executable workflow, is a directed acyclic graph that \dfp{} automatically constructs from the abstract workflow taking into account the selected mapping specified by the user. The concrete workflow is the actual workflow executed by the compute infrastructure. 

\item \textit{Grouping} governs how processing elements (PEs) communicate during input connections in \dfp{}. It offers a range of grouping choices available, each with distinct behaviors. For instance, \texttt{group-by} operates akin to ‘MapReduce,’ directing data units with matching values (e.g `state' in Figure~\ref{fig:wfsent}) in the specified element to the same PE instance. Employing these grouping strategies bolsters both data distribution and the efficiency of communication within the workflow.


\item \textit{Stateless \& Stateful}: 
In the context of \dfp{}, processing elements (PEs) can exhibit either stateless or stateful behavior. A stateless PE relies solely on its current input for operation, while a stateful PE retains information from previous inputs to influence subsequent outputs. 

\item \textit{Workload Allocation}:  Originally, \dfp{} supported only static deployment, distributing workloads without considering workflow features. In our previous work~\cite{liang2022scalable}, we introduced \dyn{} for adaptive resource allocation to PEs without halting their execution, addressing data-rate and workload variations. Nonetheless, \dyn{} exclusively manages stateless PEs and lacks support for grouping. Further details regarding \dyn{} are elaborated in Section~\ref{sec:adaptive}.

\end{itemize}

Creating \dfp{} workflows entails user-designed PEs and connections within graphs. PEs are defined with Python classes and linked by specifying inputs and outputs. Upon composition, the (abstract) workflow forms a DAG, where nodes signify PEs and edges depict data flow. Users then choose a mapping to execute the (abstract) workflow on a computing platform, with \dfp{} automatically generating the suitable concrete workflow. 




To illustrate \dfp{}'s mapping mechanism,  we are going to examine the abstract workflow and concrete workflow shown in Figure~\ref{fig:dfp_framework}. In this example, the abstract workflow is mapped using \multiprocessing{} mapping across 12 cores, with the first PE exclusively assigned to a single process. Subsequently, each of the remaining PEs is allocated 3 ($\left \lfloor{\frac{12-1}{3}}\right \rfloor $) instances, leaving 2 cores unutilized (as shown in the concrete workflow). This inefficient allocation, leading to two idle cores, prompted our exploration into the \hybrid{} and \auto{} optimizations explained in section~\ref{sec:proposed}.


\subsection{Adaptive Optimizations for \dfp{}}
\label{sec:adaptive}

In~\cite{liang2020adaptive}, we pioneered several optimization for \dfp{}, leading to the inclusion of an optimization module within the \dfp{} framework. The module includes two static optimizations and one dynamic optimization. The static methods, namely, \naive{} and \stage{}, are designed to ensure a compact PE allocation to minimize communication costs while maintaining workflow balance. Specifically, \naive{} consolidates all interconnected PEs whose communication times surpass their execution times by analyzing execution logs, while \stage{} clusters operations that do not require data shuffling based on the abstract workflow.

\begin{figure}[h!]
    \centering
    \includegraphics[width=0.8\linewidth]{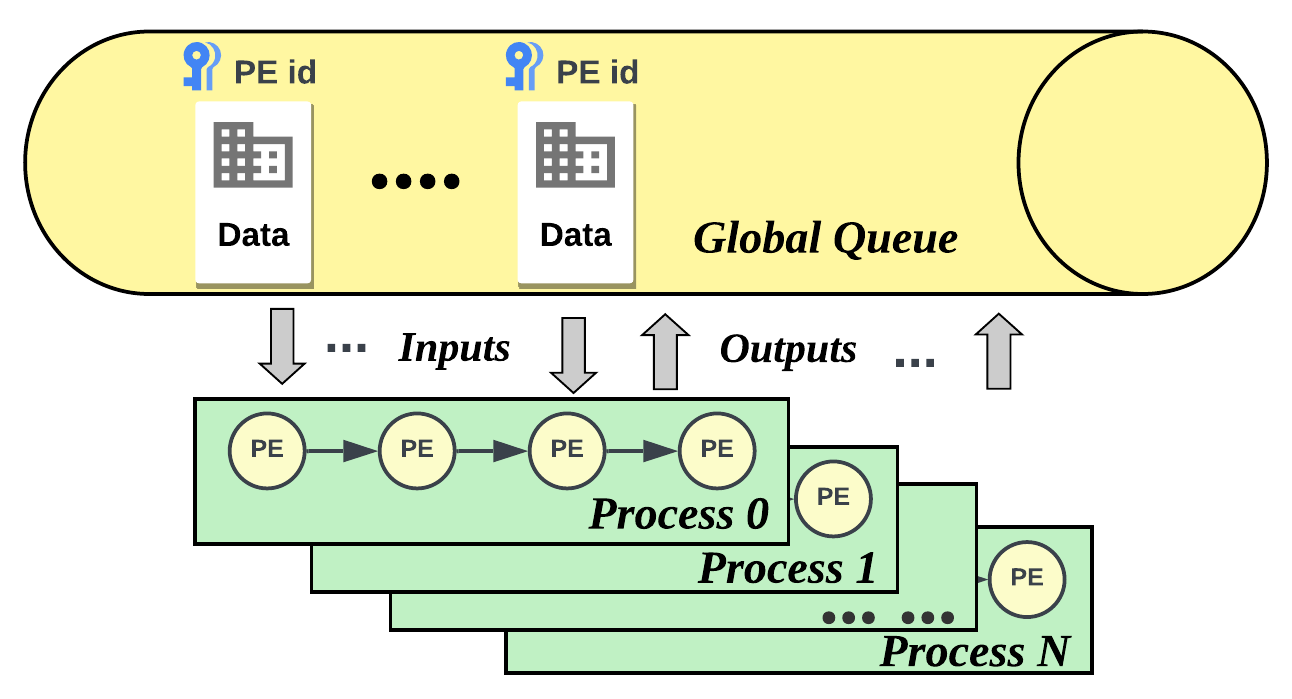}
    \caption{\dfp{}'s \dyn{} mapping.  Processes retrieve PEs dynamically from `Global Queue' and return results.}
    \label{fig:dfp_dyn}
\end{figure}

Dynamic optimization, in contrast, targets the execution phase. Unlike static optimizations that assign PEs to distinct processes upfront, the dynamic approach allocates the entire workflow graph to all processes, without predefined tasks. This shift transforms the fixed one-to-one mapping into a dynamic PE-Process mode, where task execution depends on both the PE ID and the process. Processes possess an abstract workflow map, fetching tasks and data from a global queue, referencing their maps for operations, and returning results to the global queue (as depicted in Figure~\ref{fig:dfp_dyn}). Although this shift does not affect \dfp{} end-users, it holds vital implications for developers seeking to advance the framework.



Though the optimization module functions independently of the mapping module, the ideal scenario is for the optimization methods to be compatible with various mapping techniques after certain modifications. While static optimizations seamlessly fit with different mappings since they optimize the abstract workflow prior to actual mapping, dynamic optimizations depend on the capabilities and design of the chosen mapping method. For instance, \dyn{} is ineffective with \simple{} mapping, where tasks are executed sequentially. MPI was primarily designed for SIMD (Single Instruction, Multiple Data) parallel processing, emphasizing efficient and predictable communication patterns among processes. However, this structure is at odds with the adaptive and unpredictable communication inherent to \dyn{}. Traditional MPI lacks support for a queue-based system crucial for dynamic task assignments. While \dyn{} involves regular inter-process communication for task balancing, MPI's protocols aim to minimize such communications for efficiency. Consequently, the \mpi{} mapping is not  suited for the demands of \dyn{}. 


The existing dynamic optimization of \dfp{} encounters challenges when adapted to novel computational demands, particularly in handling stateful applications. These applications necessitate maintaining consistent states across tasks, a task that \dfp{}'s random task and data selection approach struggles with. As a solution, this work introduces a hybrid strategy (see Section~\ref{sec:hybrid}) to effectively address this limitation and enable seamless handling of stateful applications.


\subsection{Redis}
\label{sec:background_redis}
Redis~\cite{eddelbuettel2022brief} is as an open-source, in-memory data structure store, renowned for its efficiency in managing a diverse array of data tasks. Supporting various data structures such as strings, lists, sets, and hashes, Redis has found widespread application in caching, real-time analytics, and messaging systems due to its rapid data access and low latency. Its minimalist yet high-performance design philosophy enables it to handle substantial data volumes seamlessly. Notably, Redis offers advanced functionalities like replication, clustering, and pub/sub messaging, bolstering its adaptability and robustness for modern data-intensive applications. 

This work delves into Redis's potential (see Section~\ref{sec:redis}) to amplify the performance and resource management of \dfp{} workflows. Particularly, the integration of Redis Stream\footnote{\url{https://redis.io/docs/data-types/streams/}}, a novel data type introduced in Redis 5.0, emerges as a focal point. Redis Stream empowers Redis with dynamic capabilities, enabling streamlined management of message and event streams. This fosters collaborative message consumption by multiple clients from a unified stream. With distinctive message entries and real-time data processing, Redis Stream seamlessly aligns with \dfp{}'s \dyn{}  objectives.

\subsection{Related Work}

Within this subsection, we delve into relevant literature, with a particular focus on two key dimensions of optimization: auto-scaling and the management of stateful applications.

\subsubsection{Auto-scaling}

Auto-scaling mechanisms find application in various domains: (1) Cloud service providers\cite{radhika2021review} like AWS, Google Cloud, and Azure offer auto-scaling for efficient resource allocation of VM instances while ensuring optimal user experience; (2) Big Data processing\cite{thonglek2021auto, alkatheri2019comparative}: Apache Spark and Apache Flink dynamically allocate resources during runtime; (3) Databases\cite{georgiou2019towards}: Amazon Aurora and Azure Cosmos DB leverage auto-scaling to maintain peak performance amidst varying workloads; (4) Workflow systems: Tools like Celery\footnote{\url{https://docs.celeryq.dev/en/stable/}} possess auto-scaling capabilities, extendable with external tools. While \dfp{} and Celery share the domain of workflow systems, their distinct objectives lead to different considerations when integrating auto-scaling. \dfp{} targets data-flow applications, while Celery caters to task-based ones.

\subsubsection{Stateful Applications}

Stateful computing in distributed computing remains a popular topic. There are multiple areas that concern statefulness: (1) Traditional relational databases such as MySQL and PostgreSQL support states across transactions.  (2) NoSQL databases such as MongoDB \cite{gupta2022optimizing} support states across nodes, ubiquitous in modern web and mobile application demands. (3) Middleware and Message Brokers: systems such as RabbitMQ maintain stateful information about messages, ensuring reliable and ordered message delivery. (4) Stream processing frameworks: Checkpointing is a prevalent strategy for state management. Global snapshot is a state-of-the-art periodic checkpointing solution which captures the holistic state of execution. For instance, Apache Flink has introduced a sophisticated distributed asynchronous snapshot mechanism \cite{de2021distributed}. This ensures a low-cost state management mechanism; however, it depends on Apache Flink’s distinct data flow, which guarantees the ordering. Another check-pointing method for state management is the localized state used by Apache Storm Trident \cite{jain2017mastering}. Both checkpointing methods require ordering, a guarantee that dynamic \dfp{} cannot provide.

\section{Mappings and Optimizations}
\label{sec:proposed}

In this work, we have directed our efforts towards two primary objectives: leveraging the potential of the Redis framework to integrate the \dyn{} optimization and introducing an innovative auto-scaling strategy. The upcoming subsections explain in detail the proposed techniques.

\subsection{Redis Mappings}
\label{sec:redis}

The incorporation of Redis into \dfp{} brings the potential for optimizing data-intensive workflows, as discussed in Section~\ref{sec:background_redis}. Redis, recognized for its effective data management, introduces transformative features through Redis Stream. This real-time sequence maintenance aligns with the \dfp{} dynamic optimizations.

In our pursuit of integrating \dfp{} with Redis, we initially introduced the \textit{dynamic Redis} mapping. Yet, to enhance support for groupings and stateful applications within dynamic optimization, we have introduced the \textit{hybrid Redis} mapping. This addition further extends \dfp{}'s capabilities to accommodate a broader range of optimization scenarios

\subsubsection{Dynamic Redis Mapping}
\label{sec:dyn_redis}

The \textit{dynamic Redis} mapping draws inspiration from the original \dyn{} for \multiprocessing{} mapping. In this adaptation, the multiprocessing queue is replaced with the powerful Redis stream, seamlessly incorporating Redis's inherent features. This alteration is exemplified in Figure~\ref{fig:dfp_dyn}, where the previous `Global Queue' utilizing a multiprocessing queue has been replaced by the Redis stream mechanism for this new mapping.

\subsubsection{Hybrid Redis Mapping}
\label{sec:hybrid}

\begin{figure}[h!]
    \centering
    \includegraphics[width=1\linewidth]{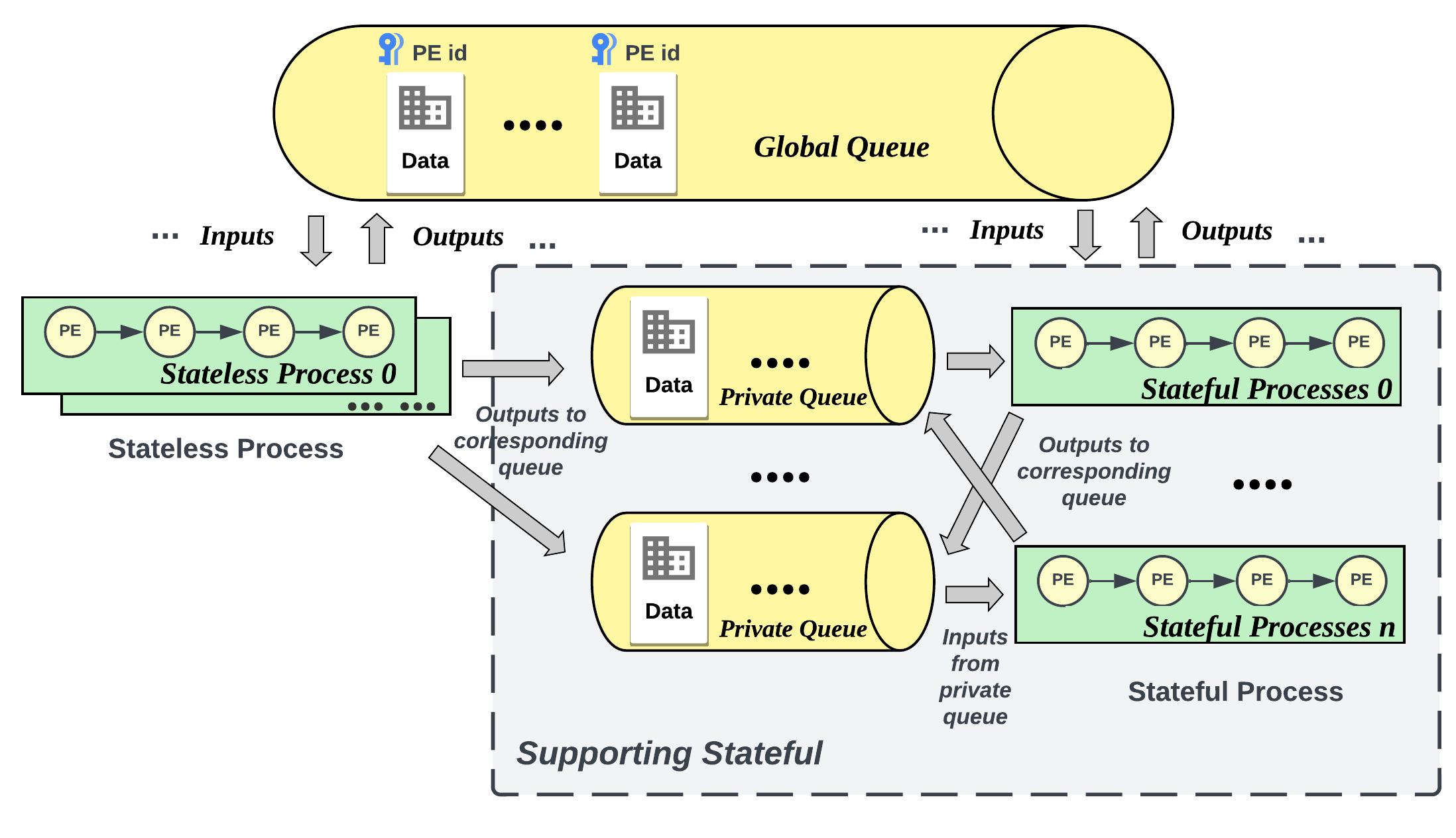}
    \caption{\dfp{}'s \textit{hybrid Redis} mapping. Processes categorized into stateless and stateful manage corresponding PEs, optimizing workflow execution. Stateful processes maintain dedicated private queues.}
    \label{fig:d4p_hybrid}
\end{figure}

To address the requirements of stateful applications and dynamic optimization, we introduce the \textit{hybrid Redis} mapping (illustrated in Figure~\ref{fig:d4p_hybrid}). This novel mapping strategy is designed to efficiently handle both stateless and stateful tasks. Key to its operation is the direct mapping of stateful PE instances to dedicated processes, ensuring the maintenance of local states and private task input queues (refereed as `Private Queues' in Figure~\ref{fig:d4p_hybrid} ). Meanwhile, outputs from these stateful PE instances can be routed to different queues based on connections, eliminating the need for continuous state synchronization and enhancing performance in comparison to traditional global state management approaches.


In various stateful applications, tasks can exhibit either a stateful or stateless nature. To maximize the advantages offered by the \dyn{} mapping strategy, the \hybrid{} mapping introduces a mechanism where stateless PE instances are assigned to the available processes that are not dedicated to stateful tasks. This allocation is determined by calculating $N - \text{number of stateful PE instances}$, where $N$ represents the total number of available processes. These stateless processes function similarly to the conventional dynamic methods, acquiring inputs from the `Global queue' and returning completed tasks to it. However, a subtle distinction arises: these stateless tasks possess the additional capability of depositing their outputs into private queues specifically designated for stateful tasks, enhancing the efficiency of the overall workflow execution.

\subsection{Auto-scaling Optimizations}
\label{sec:auto-scaling}


The \auto{} optimization addresses the efficient allocation of computational resources in response to the varying workload. \auto{} makes the system responsible for workload spikes, and reduces resource wastage during low workload periods, thereby improving the efficiency of \dfp{}. We implement \auto{}  optimization for \multiprocessing{} and \textit{dynamic Redis} (see  Section~\ref{sec:dyn_redis}) mappings for handling stateless workflow, namely \textit{dynamic auto-scaling Multiprocessing} and \textit{dynamic auto-scaling Redis}.

\begin{figure}
    \centering
    \includegraphics[width=1\linewidth]{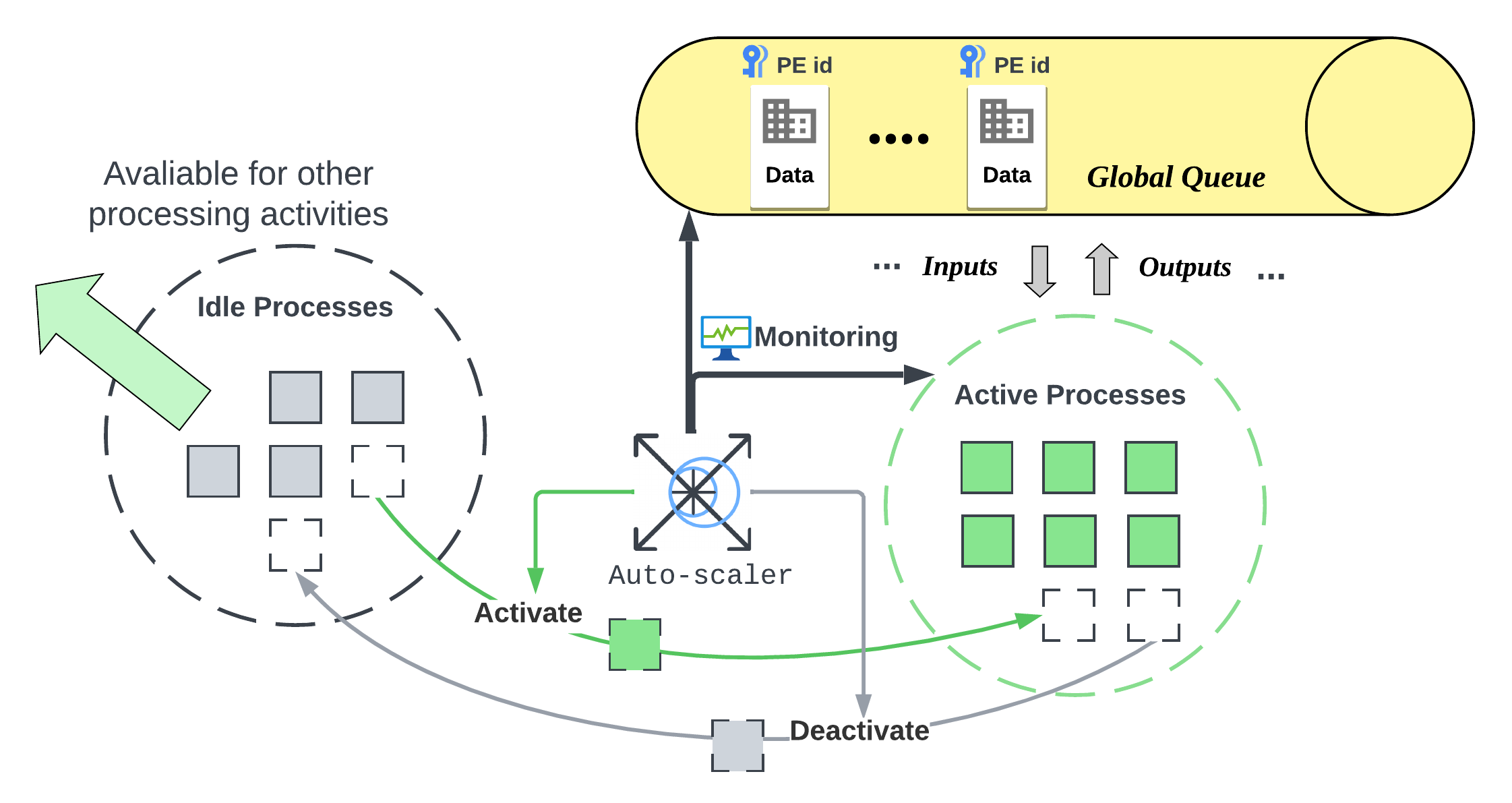}
    \caption{\dfp{}'s \auto{}. Idle processes conserve energy and can be reactivated when workload surges.}
    \label{fig:dfp_auto}
\end{figure}

Within the framework of \dfp{}, \auto{} extends the capabilities of \dyn{}, as depicted in Figure~\ref{fig:dfp_auto}. This enhancement introduces two processor statuses: active and idle. Active processes are allocated to the abstract workflow and actively participate in task execution. They retrieve PE IDs and associated data from the queue, process them, and return the results, similar to the original \dyn{} behavior.


However, a notable distinction arises when the queue experiences a reduced workload. In the conventional \dyn{} framework (introduced in Section~\ref{sec:adaptive}), certain processes continuously monitor the queue, anticipating new tasks. This operation, however, proves to be both resource-intensive and redundant. \auto{} introduces an optimized approach to address this issue. Processes without immediate tasks are transitioned into an idle state, operating in a low-energy standby mode. This efficient mechanism not only curbs energy consumption but also offers the potential for reallocating these idle processes elsewhere. When a surge in workload is detected, \auto{} can swiftly activate these dormant processes. Conversely, during periods of low demand, surplus processes are deactivated and returned to an idle state. This dynamic activation and deactivation process is orchestrated by the \textit{auto-scaler}, working in tandem with a monitoring framework and various auto-scaling strategies.

\subsubsection{Auto-scaler}


Algorithm~\ref{alg:autoscaler} shows how the \textit{auto-scaler} works. It sets parameters like the maximum pool size and workload threshold upon initialisation; by default, the $active\_size$ is half of the maximum of total processes ($max\_ pool\_size$). The configurable $workload\_threshold$ is used in auto-scaling strategies. The \texttt{shrink} and \texttt{grow} methods decrease and increase the active processes, respectively, and are controlled by the central logic methods \texttt{auto\_scale}. It implements auto-scaling strategies (introduced in the Section~\ref{subsubset:auto-scaling_strategy}), which monitor the system state, and control the resource adjustments. Tasks are dispatched through the \texttt{start} method and finished by the \texttt{done} methods; those two methods are also responsible for updating the active count denoted as $active\_count$ used to avoid over-use of processes. The entry method \texttt{process} continually evaluates and scales resources before activating processes by calling \texttt{start}. Overall, the \textit{auto-scaler} can schedule the resources dynamically, ensuring efficiency and responsiveness.

\begin{algorithm}
\caption{Auto\_scaler for Dynamic Optimization}
\label{alg:autoscaler}
\begin{algorithmic}[1]

\State \textbf{Class} Auto\_scaler:

    \State \textbf{\textit{Parameters}} : $max\_pool\_size$, $pool$, $threshold$, $queue$, $active\_size$, $active\_count$
    \\
    \Procedure{Constructor}{$\textbf{\textit{Parameters}}$}
        \State Initialize member parameters
        \State By default $active\_size$ $\gets$  $max\_pool\_size$/2
    \EndProcedure
    \\
    \Procedure{shrink}{$size\_to\_shrink$}
        \State Decrease $active\_size$ by $size\_to\_shrink$ (with a minimum of 1)
    \EndProcedure
    \\
    \Procedure{grow}{$size\_to\_grow$}
        \State Increase $active\_size$ by $size\_to\_grow$ (with a maximum of $max\_pool\_size$)
    \EndProcedure
    \\
    \Procedure{is\_terminiated}{}
        \State \textbf{Return} \textit{BOOL} depending on \textbf{Termination Methods} 
    \EndProcedure
    \\
    \Procedure{auto\_scale}{}
        \State $curr\_states \gets$ \textbf{Monitor}
        \If{$curr\_states$ > $threshold$}
            \State \Call{grow}{1}
        \Else
            \State \Call{shrink}{1}
        \EndIf
    \EndProcedure
    \\
    \Procedure{start}{$func, args$}
        \While{$active\_count$ >= $active\_size$}
            \State Wait
        \EndWhile
        \State Increment $active\_count$
        \State \Return $Pool$.apply\_async($func$, $args$, callback=\Call{done})
    \EndProcedure
    \\
    \Procedure{done}{$result$}
        \State Decrement $active\_count$
    \EndProcedure
    \\
    \Procedure{process}{$graph$}
        \While{True}
            \State \Call{auto\_scale}{}
            \If{\Call{is\_terminiated}}
                \State Get results from $results$
                \State \textbf{Return}
            \Else
                \State $cp\_graph \gets$ DeepCopy($graph$)
                \State Initialize worker's $args$ with $queue$ and $cp\_graph$ 
                \State \Call{start}{worker.process, $args$}
            \EndIf
        \EndWhile
    \EndProcedure

    \State \textbf{end Class}
\end{algorithmic}
\end{algorithm}

\subsubsection{Auto-scaling and Monitoring Strategy}
\label{subsubset:auto-scaling_strategy}

Auto-scaling is not just about dynamically reallocating resources; striking the right balance between performance and efficiency is also the crux of chosen auto-scaling strategies. These strategies have two important decisions: `when to scale' and `how to scale'. The former is about metrics from the monitoring framework, while the latter determines the magnitude of scaling. For instance, a question might be whether we should rapidly scale up when we observe a task burst. In this work, we adopt a simple incremental approach: incrementing the active size by 1 or -1. Given \textit{dynamic auto-scaling Multiprocessing} and \textit{dynamic auto-scaling Redis} have different monitoring frameworks, we use a different strategy for each:

\begin{itemize}
    \item \textit{dynamic auto-scaling Multiprocessing}: This approach employs queue size to gauge the current workload. When the queue size increases compared to the previous state, indicating higher task volume, additional processes are activated. Conversely, processes are deactivated during reduced workload, while a minimum threshold prevents unnecessary scaling during low demand.

    \item \textit{dynamic auto-scaling Redis}: Here we utilize Redis's consumer group's average idle time as a metric. Unlike queue size, idle time directly reflects process states. We employ a threshold for the average idle time of active processes. If a process's idle time exceeds the time needed for reactivation and redeployment, it is logically deactivated. The reactivation time is influenced by computational resources and the specific workflow, requiring proper configuration for practical use
\end{itemize}

The experimental results in Section~\ref{sec:exp} show that these preliminary strategies are effective. While they currently serve our purpose, we will dive deeper to refine and optimize them in future work.

\subsubsection{Termination Strategy}


As mentioned earlier, with the shift to dynamic optimization, how tasks are executed and terminated significantly changes. Instead of the static pre-assignment of PEs to specific processes, dynamic optimization assigns the entire workflow graph to all processes and takes a task from the queue without order. Such changes results in the failure of traditional static termination methods.  In the static context, the \textit{"poison pills"} termination method was employed, where the source PE would signal the end of its input to all subsequent instances. However, in the dynamic setting, the task processing order is not reserved; it is based on availability rather than any specific order in the abstract workflow. This makes the \textit{"poison pills"} method ineffective, as it can unexpectedly halt processes and leave tasks unfinished in the queue.  To solve this, the native dynamic approach relies on checking the emptiness of the global queue for termination. While this method is generally effective, it is not foolproof and could lead to unexpected exits in some extreme cases. Additionally, constant checks from all processes on the queue's status could be inefficient. 


We use a retry mechanism combined with \textit{"poison pills"} to mitigate these challenges. If the queue appears empty, processes will wait for a configurable threshold duration and retry a specified number of times before deciding on termination. Once a process determines to stop based on these parameters, it broadcasts \textit{"poison pills"} to other processes, speeding up their termination check, thereby reducing overall waiting time.


\section{Use Cases}
\label{sec:usecase}

In this section, we introduce three real scientific \dfp{} workflows, including two stateless workflows and one stateful workflow. These workflows will be used in the experiment to evaluate our proposed mappings and optimizations.



\subsection{\wfint{}}



\begin{figure}[h!]
    \centering
    \includegraphics[width=0.85\linewidth]{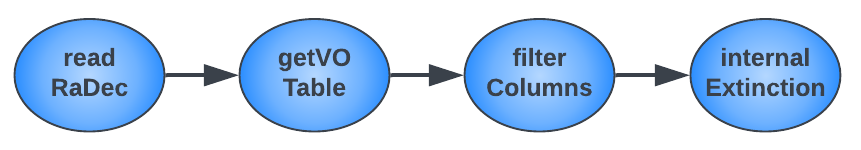}
    \caption{\wfint{} workflow.}
    \label{fig:wfint}
\end{figure}

The \wfint{} workflow has been implemented to calculate the extinction metric within galaxies, which is a significant property in astrophysics. This property reflects the dust extinction of the internal galaxies and is used for measuring optical luminosity. The workflow contains four PEs, as shown in Figure~\ref{fig:wfint}. The first PE, \texttt{read RaDec}, reads the coordinator data for galaxies in an input file. Then, these data are used in \texttt{getVO Table} to download the VOTables. These VOTables in \texttt{filter Columns} are filtered by specified columns used in the internal extinction computation. The last PE (\texttt{internal Extinction}) will perform the computation. It is important to note that all PEs here are stateless. 

To introduce variability in the workload, we adjusted the \texttt{read RaDec} PE. For a standard workload (denoted as 1X), it reads data for 100 galaxies. This reading scales to 300 galaxies for 3X, 500 for 5X, and 1000 for 10X. In addition to varying the stream length, we also varied the PE's workload. By using a random sleep time sampled from a beta(2,5) distribution, we added delays ranging from 0 to 1 second within the \texttt{getVO Table} PE and \texttt{filter Columns} PE. This modification is labeled as "heavy". Therefore, the experiment based on \wfint{} now has both standard and heavy workloads varying from 1X to 10X.

\subsection{\wfcorr{}}

\begin{figure*}
    \centering
    \includegraphics[width=1\linewidth]{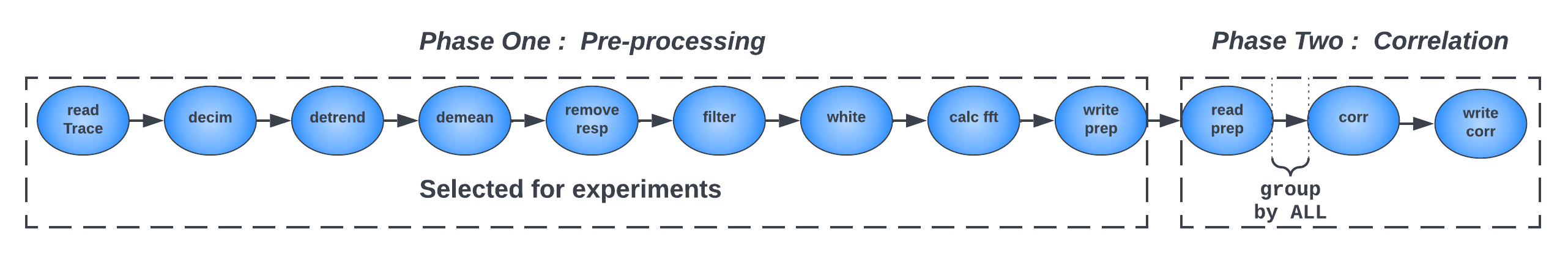}
    \caption{\wfcorr{} workflow.}
    \label{fig:wfcorr}
\end{figure*}

The \wfcorr{} workflow is engineered to monitor and analyze the vast geological waveform data gathered from FDSN\footnote{\url{https://www.fdsn.org}} stations. Its primary goal is to evaluate and predict the risk and likelihood of volcanic eruptions and earthquakes in real-time. The original workflow consists of two phases (in Figure~\ref{fig:wfcorr}): the initial stage involves pre-processing data collected from stations, and the second phase deals with reading the pre-processed data and executing the cross-correlation computations.


Notably, the second phase has a \textit{grouping} mechanism. Given that the \auto{} cannot handle stateful applications, we want to focus on the first phase of our experiments (all PEs in the first phase are stateless). For testing the \hybrid{}, we selected another representative workflow. The first phase comprises nine interconnected PEs: the initial PE reads the data, the intermediate PEs process the raw data, and the final PE writes the data to disk. There are more imbalanced workloads among PEs; for example, the intermediate PEs only do calculations in main memory, but the last PE writes data into the disk, which involves IO operations. A detailed description of the setup can be found in~\cite{liang2020adaptive, liang2022scalable}.

\subsection{\wfsent{}}


This workflow analyses newspaper articles data by implementing a sentiment analysis of the news\footnote{\url{https://github.com/NoPuzzle/dispel4py_autoscaling/tree/main/dispel4py/examples/article_sentiment_analysis}}. This workflow applies two distinct sentiment analyses on articles, subsequently aggregating these sentiment scores based on the published location. Such workflow complicates the graph by having various types of \textit{grouping} compared with the first two workflows.  The news articles employed the source data from public \textit{News Articles} datasets on Kaggle\footnote{\url{https://www.kaggle.com/datasets/asad1m9a9h6mood/news-articles}}. 



\begin{figure}[h!]
    \centering
    \includegraphics[width=1\linewidth]{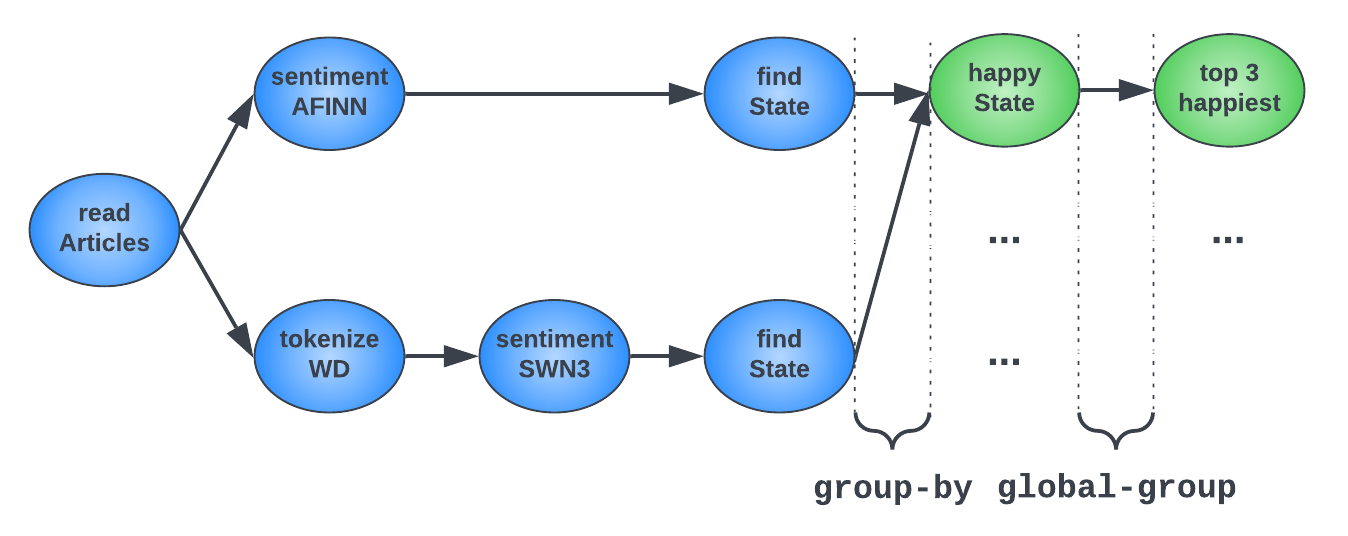}
    \caption{\wfsent{} workflow.}
    \label{fig:wfsent}
\end{figure}

The workflow shown in Figure~\ref{fig:wfsent} unfolds as follows. The initial PE (\texttt{read Articles}) sequentially reads and parses articles from input files. Each parsed article then undergoes dual processing by two downstream PEs. The \texttt{sentiment AFINN} PE computes sentiment scores utilizing the AFINN lexicon\footnote{\url{https://github.com/fnielsen/afinn}}, while concurrently, \texttt{tokenize WD} and \texttt{sentiment SWN3} PEs tokenize the articles and derive sentiment using the SWN3 lexicon \cite{abdul2012toward}. Post-processing, the data from both pathways convene within their respective \texttt{find State} - \texttt{happy State} - \texttt{top 3 happiest} sequence. These three PEs identify, group, and display the top three happiest locations with their scores.


Stateful PEs play a crucial role in this workflow. For instance, the \texttt{happy State} PE is strategically distributed across four instances grouped by their `state' (\texttt{group-by}), while the \texttt{top 3 happiest} PE is confined to one instance under the \texttt{global} grouping~\footnote{This grouping enforces all instances of the previous PE (characterized by the `happy state') are directed towards a singular instance of the same PE (representing the `top 3 happiest')}. In contrast, other PEs lack these constraints and are classified as stateless. By blending stateless and stateful PEs, this workflow stands as an ideal testbed to explore the behavior of enhanced dynamic deployment within the realm of a real stateful application.

\begin{figure}
    \centering
    \includegraphics[width=0.95\linewidth]{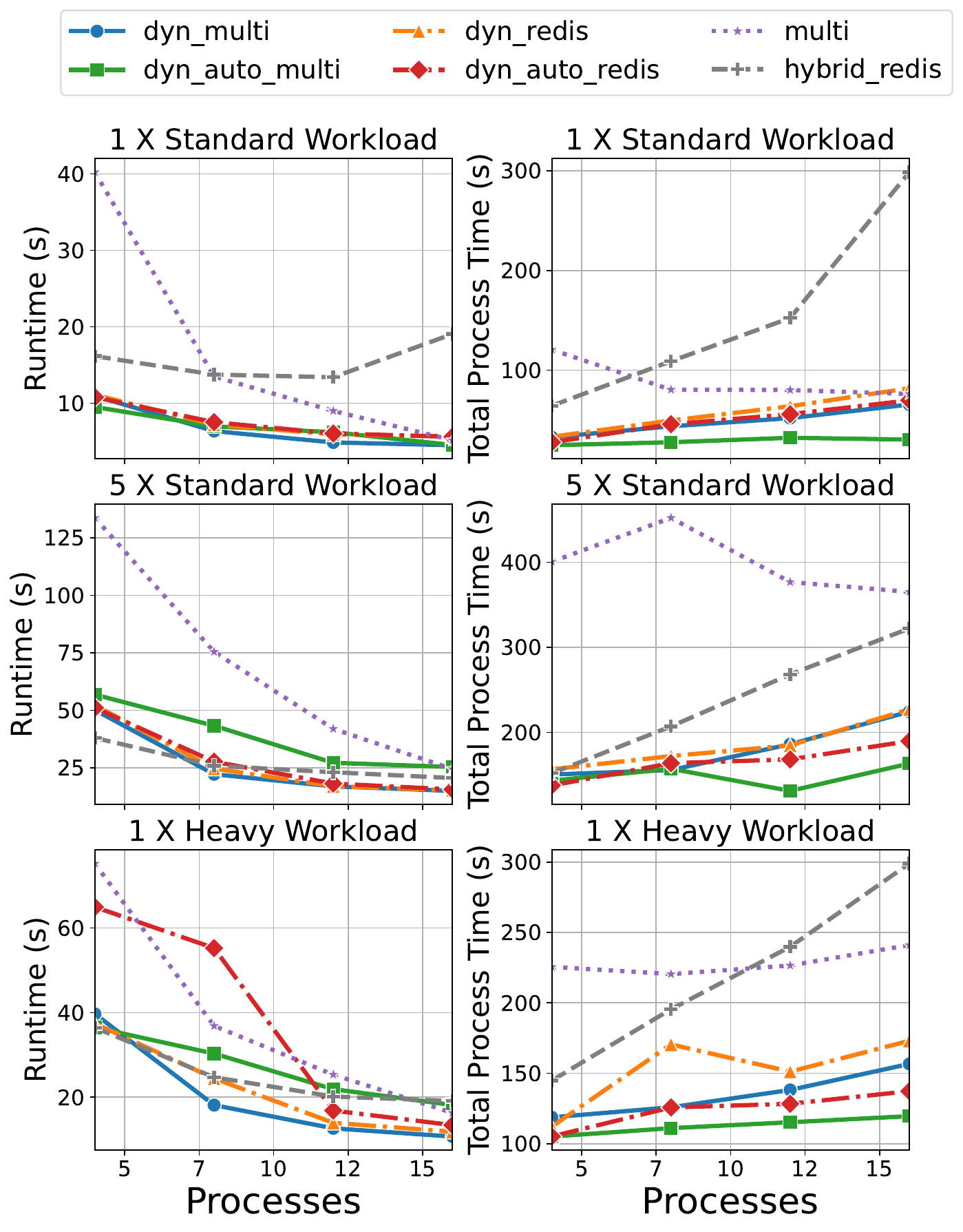}
    \caption{Workload Performance for \wfint{} using the \server{} with up to 16 processes.}
    \label{fig:int_server}
\end{figure}

\begin{figure}
    \centering
    \includegraphics[width=0.95\linewidth]{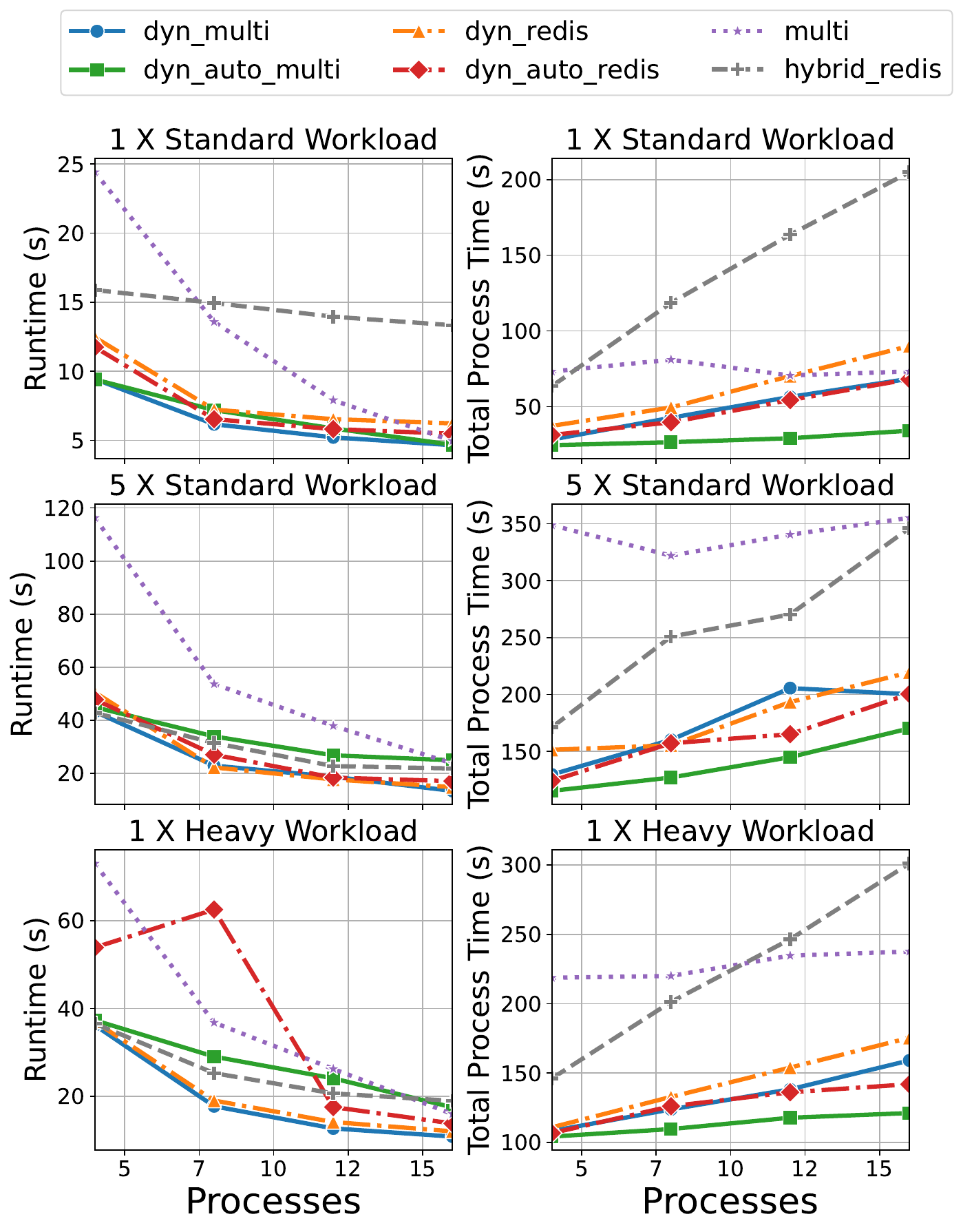}
    \caption{Workload Performance for \wfint{} using the \cloud{} with up to 16 processes.}
    \label{fig:int_cloud}
\end{figure}

\begin{figure}
    \centering
    \includegraphics[width=0.95\linewidth]{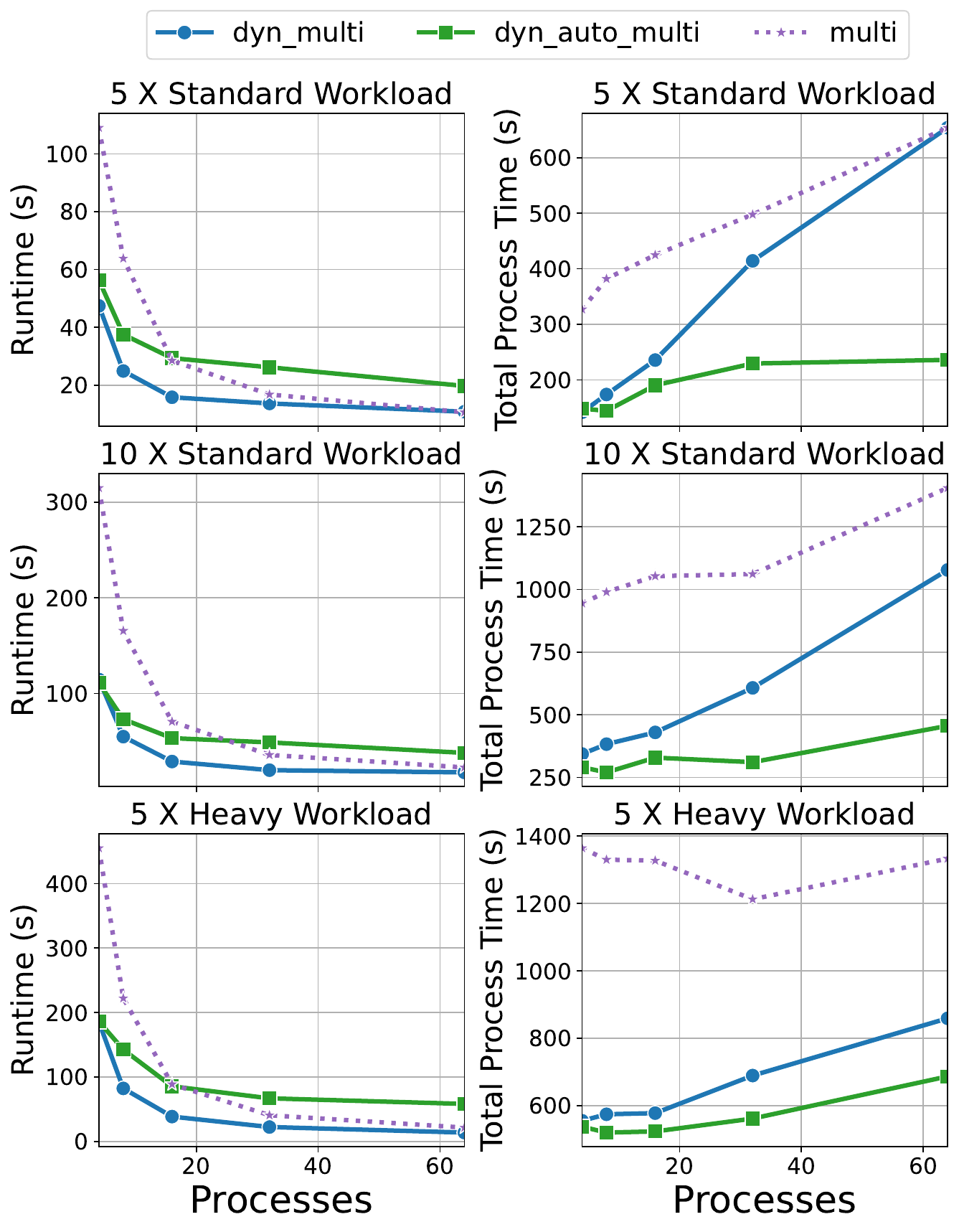}
    \caption{Workload Performance for \wfint{} using the \hpc{} with up to 64 processes.}
    \label{fig:int_hpc}
\end{figure}

\section{Evaluations}
\label{sec:exp}

In this section, we conduct a comparative experimental evaluation of the workflows outlined in Section~\ref{sec:usecase}. Our primary focus lies in assessing the performance and efficiency of the mappings and optimization techniques proposed in Section~\ref{sec:proposed}. We introduce abbreviations and provide key insights into the evaluated techniques:

\begin{itemize}
 
\item \multi{}: This represents the native \multiprocessing{} mapping~\cite{filgueira2015dispel4py}. this approach statically assigns PE instances to processes as detailed in the background Section~\ref{sec:d4py_background}. Benefiting from its inherent state maintenance capabilities, \multi{} can effectively manage both stateful and stateless applications, establishing itself as an appropriate baseline for all experimentation.

\item \dynmulti{}: This denotes the \textit{dynamic Multiprocessing} mapping~\cite{liang2022scalable} introduced in the background Section~\ref{sec:adaptive}~. It enables processes to dynamically share the workload on-the-fly. Being built upon the \multiprocessing{} mapping, \dynmulti{} serves as a baseline for comparing the \auto{} optimization based on the same mapping.

\item \automulti{}: This signifies the new \textit{dynamic auto-scaling Multiprocessing} mapping presented in Section~\ref{subsubset:auto-scaling_strategy}. It will be compared primarily with \dynmulti{}, aiming to evaluate the efficiency gains achieved by incorporating \auto{}.

\item \dynredis{}: This refers to the new \textit{dynamic Redis} mapping explained in Section~\ref{sec:dyn_redis}. The performance of \dynredis{} is horizontally compared with \dynmulti{}, as both employ the \dyn{} optimization.

\item \autoredis{}: This stands for the new \textit{dynamic auto-scaling Redis} mapping introduced in Section~\ref{subsubset:auto-scaling_strategy}. \autoredis{} is mainly compared with \dynredis{} to evaluate \auto{} in a vertical comparison, given their shared mapping foundation. Additionally, it is compared with \automulti{} for a horizontal assessment over different mappings.

\item \hyredis{}: This denotes the new \textit{hybrid Redis} mapping explained in Section~\ref{sec:hybrid}, exclusively designed to support stateful applications. The performance of \hyredis{} will be comparatively evaluated with \multi{}.
\end{itemize}

\subsection{Experiment Setup}

\subsubsection{Platform}


We have conducted our experiments on multiple infrastructures/platforms, each with its distinct configurations:

\begin{itemize}
    \item \server{}: This is a virtual server for research groups supported by Department of Computing, Imperial College London \footnote{\url{https://www.imperial.ac.uk/admin-services/ict/self-service/research-support/private-cloud/}}. It consists of 16-core with Intel E5-2690@2.60GHz processor paired with 64GB of RAM and runs on Ubuntu 14.04. We denote it as \server{} for short. It runs all three workflows on various processes: 4, 8, 12, and 16.

    \item \cloud{}: This server is provided by Google Cloud Platform \footnote{\url{https://cloud.google.com/}}. It has 4 Intel(R) Xeon(R) CPU @ 2.20GHz (8vCPUs), and 16GB of RAM running Ubuntu 20.04. For our experiments, we refer this server to \cloud{}, and its configuration for running the experiment is identical to \server{}.

    \item \hpc{}: The HPC cluster we used is provided by HPC at Imperial \footnote{\url{https://www.imperial.ac.uk/computational-methods/hpc/}}. The HPC servers provide multi-job classes, and we use the \textit{short} class in which the CPU and OS mode are Intel E5-2680 v3 @ 2.50GHz and Centos 8, respectively. We request up to 64 CPUs and 64 GB memory. Since Redis cannot be deployed on the HPC, no mapping is based on Redis running on HPC. We employed 4, 8, 16, 32, and 64 CPUs for other experiments. 
\end{itemize}


\subsubsection{Metrics}

We use two main metrics for evaluation: runtime (\rtime{}) and total process time (\ptime{}). \rtime{} represents the real-world execution time, while \ptime{} accounts for all active process durations, reflecting overall efficiency. Typically, fewer processes lead to shorter \ptime{} due to reduced synchronization overhead, but may extend \rtime{}. Our results are presented as \rtime{} and \ptime{} pairs for a comprehensive view.

Moreover, we calculate \rtime{} and \ptime{} ratios between methods to provide intuitive insights. For example, a \rtime{} ratio below 1 (shown in Table~\ref{tab:int}) between \automulti{} and \dynmulti{} implies that \automulti{} completes tasks faster. Similarly, a \ptime{} ratio below 1 suggests \automulti{} is more efficient than \dynmulti{}. To maintain consistency, we only include our proposed optimizations (\auto{} or \hyredis{}) in the numerator. If both ratios are below 1, the proposed approach excels in both performance and efficiency, although practical trade-offs may affect ideal outcomes.

\subsection{Ev. \wfint{}}


Different workloads based on \wfint{} workflow are used on multiple platforms to provide a comprehensive evaluation. On both \server{} and \cloud{}, we use three different workloads: the 1X standard workload (100 galaxies as input), the 5X standard workload (500 galaxies), and the 1X heavy workload (100 galaxies with sleep synthetically added to the PEs). On \hpc{}, which offers more cores, we deploy a heavier workload than \server{} and \cloud{}. \hpc{} will be experimented with 5X, 10X standard and 5X heavy workload.


Figure~\ref{fig:int_server} shows the performance metrics in terms of  \rtime{} (left) and \ptime{} (right) for all six techniques introduced in Section~\ref{sec:exp}. All techniques show a decreasing trend for \rtime{} with an increase in the number of processes, indicating they share good scalability in \dfp{}. With lower workloads, \auto{} techniques achieve shorter \rtime{}, as they dynamically adjust the number of activated processes, thereby lowering unnecessary synchronization costs. However, as the workload increases, particularly in heavy-workload scenarios, \auto{} technique lags slightly behind pure \dyn{} optimizations. This minor fault could be optimized with a more refined auto-scaling strategy. Apart from this, \auto{} techniques still outperform native \multi{} mapping in most cases. Regarding \ptime{}, as the number of processes increases, the \ptime{} rises due to the accumulated synchronization overhead from more active processes. Unsurprisingly, both \automulti{} and \autoredis{} increase slightly, with both excelling over their corresponding pure \dyn{} competitors. 


Notably, in the race track of versatile methods (supporting both stateless and stateful), \hyredis{} presents strong \rtime{} performance, but given we did not equip \auto{} optimization to it, \hyredis{} does not achieve the same efficiency, compared with \autoredis{}.  In terms of comparison between \multi{} and \redis{}, there is a common pattern trend across various experimental setups: both \rtime{} and \ptime{} metrics for the optimization using the \redis{} mapping are larger than those of the \multi{} mapping. The reason can be attributed to the inherent characteristics of the two mappings. Compared with \redis{}, \multi{} is designed to be lightweight, offering outstanding performance. However, \redis{} supports more features regarding monitoring, reliable messaging and robust data persistence, which render \redis{} more resource-intensive, thereby affecting its performance. 


Performance trends on the \cloud{} platform are similar to those observed on the \server{}. However, since there are only 8 cores in \cloud{}, the performance slightly dips with 12 and 16 processes compared to \server{}. Despite this, the overall trends remain consistent across different platforms, indicating the reproducibility and portability of our experiments.


Experiments on \hpc{} show how the performance metrics (both \ptime{} and \rtime{}) change with a larger scale in processes. Experiments on \hpc{} show how the performance metrics (both \ptime{} and \rtime{}) change with a larger scale in processes. The \rtime{} of all three methods based on \multi{} mapping show a quick drop when increasing the number of processes to 16, then gradually becoming steady with a slight decrease. However, the \ptime{} of \dynmulti{} and \automulti{} show a linear increase with the increase in the number of processes. In the meanwhile, the increase of \automulti{} is at a slight upward slope. This difference strongly supports the effectiveness of \auto{}, especially when a large number of processes are involved.

\subsubsection{Summary of the evaluation on \wfint{} with \rtime{} and \ptime{} ratios}


To provide an intuitive view of the results, we calculate \rtime{} and \ptime{} ratios to intuitively compare \auto{} and \dyn{} techniques: \automulti{} and \dynmulti{}, \autoredis{} and \dynredis{}. Table~\ref{tab:int} displays these ratios across various platforms, allowing analysis from different perspectives. For example, in the best \rtime{} case, \automulti{} requires only 87\% \rtime{} and 76\% \ptime{} of \dynmulti{}. Prioritizing \ptime{}, the most efficient ratio achieved is 0.46, with a 1.01 \rtime{} ratio. Overall, \auto{} techniques demonstrate efficiency by slightly extending \rtime{} while reducing \ptime{}, highlighting their trade-offs and benefits.

{
\footnotesize{
\begin{table}[]
\caption{Performance comparison based \wfint{} workflow between techniques (A and B). The \rtime{} ratio is the \rtime{}  of mapping A over the \rtime{}  of mapping B. Similarly, the \ptime{} ratio is the total process time elapsed ratio between A and B. The ratios are prioritized by the metric in the "Prioritized By" column, and the [Mean, Std] shows the average and standard deviation for all \rtime{} and \ptime{} ratios.}
\label{tab:int}
\begin{tabular}{|l|l|l|l|l|}
\hline
Platform                   & \begin{tabular}{@{}l@{}}Comparison \\ between A/B \end{tabular}                                             & Prioritized By         & Runtime Ratio              & Process Time Ratio \\ \hline
\multirow{6}{*}{\server{}} & \multirow{3}{*}{$\frac{\automulti{}}{\dynmulti{}}$} & \rtime{}        & 0.87             & 0.76               \\ \cline{3-5} 
                           &                                                     & \ptime{}        & 1.01             & 0.46               \\ \cline{3-5} 
                           &                                                     & {[}Mean, Std{]} & {[}1.39, 0.37{]} & {[}0.77, 0.15{]}   \\ \cline{2-5} 
                           & \multirow{3}{*}{$\frac{\autoredis{}}{\dynredis{}}$} & \rtime{}        & 0.97             & 0.83               \\ \cline{3-5} 
                           &                                                     & \ptime{}        & 2.26             & 0.74               \\ \cline{3-5} 
                           &                                                     & {[}Mean, Std{]} & {[}1.21, 0.39{]} & {[}0.86, 0.06{]}   \\ \hline
\multirow{6}{*}{\cloud{}}  & \multirow{3}{*}{$\frac{\automulti{}}{\dynmulti{}}$} & \rtime{}        & 1.00             & 0.87               \\ \cline{3-5} 
                           &                                                     & \ptime{}        & 1.01             & 0.50               \\ \cline{3-5} 
                           &                                                     & {[}Mean, Std{]} & {[}1.36, 0.34{]} & {[}0.77, 0.15{]}   \\ \cline{2-5} 
                           & \multirow{3}{*}{$\frac{\autoredis{}}{\dynredis{}}$} & \rtime{}        & 0.88             & 0.76               \\ \cline{3-5} 
                           &                                                     & \ptime{}        & 0.88             & 0.76               \\ \cline{3-5} 
                           &                                                     & {[}Mean, Std{]} & {[}1.26, 0.66{]} & {[}0.86, 0.08{]}   \\ \hline
\multirow{3}{*}{\hpc{}}    & \multirow{3}{*}{$\frac{\automulti{}}{\dynmulti{}}$} & \rtime{}        & 0.97             & 0.85               \\ \cline{3-5} 
                           &                                                     & \ptime{}        & 1.83             & 0.36               \\ \cline{3-5} 
                           &                                                     & {[}Mean, Std{]} & {[}1.95, 0.83{]} & {[}0.75, 0.20{]}   \\ \hline
\end{tabular}
\end{table}
}
}

\begin{figure}
     \centering
     \begin{subfigure}{1\linewidth}
         \centering
         \includegraphics[width=1\linewidth]{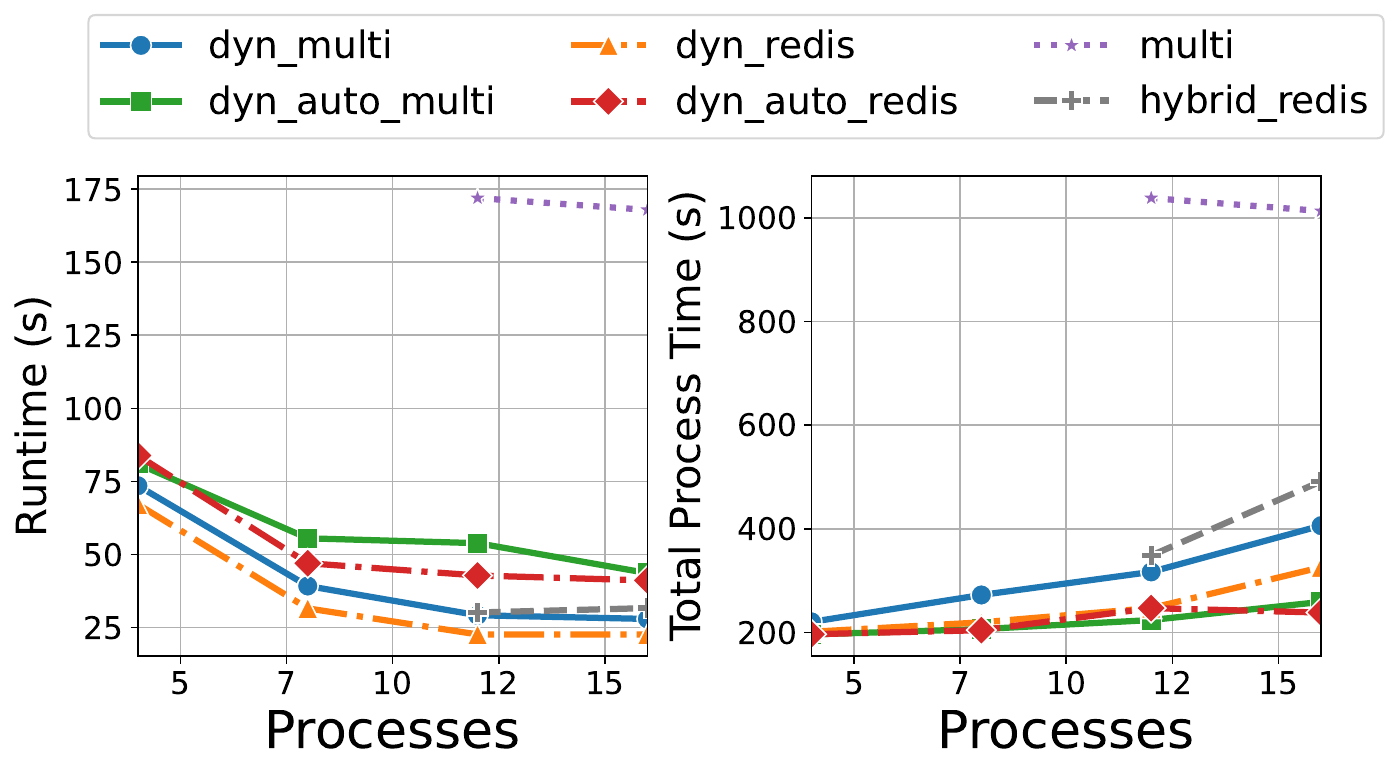}
         \caption{Workload Performance for \wfcorr{} using the \server{} with up to 16 processes.}
         \label{fig:corr_server}
    \end{subfigure}
    
    \begin{subfigure}{0.95\linewidth}
         \centering
         \includegraphics[width=1\linewidth]{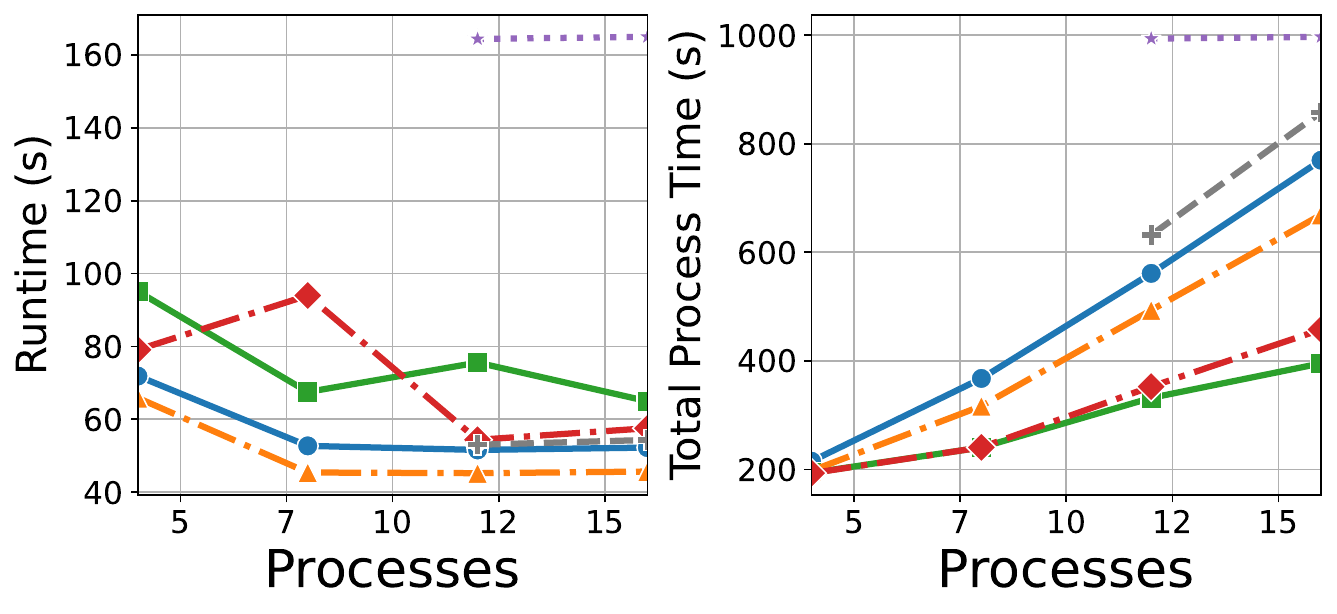}
         \caption{Workload Performance for \wfcorr{} using the \cloud{} with up to 16 processes.}
         \label{fig:corr_cloud}
    \end{subfigure}
    
    \begin{subfigure}{0.95\linewidth}
         \centering
         \includegraphics[width=1\linewidth]{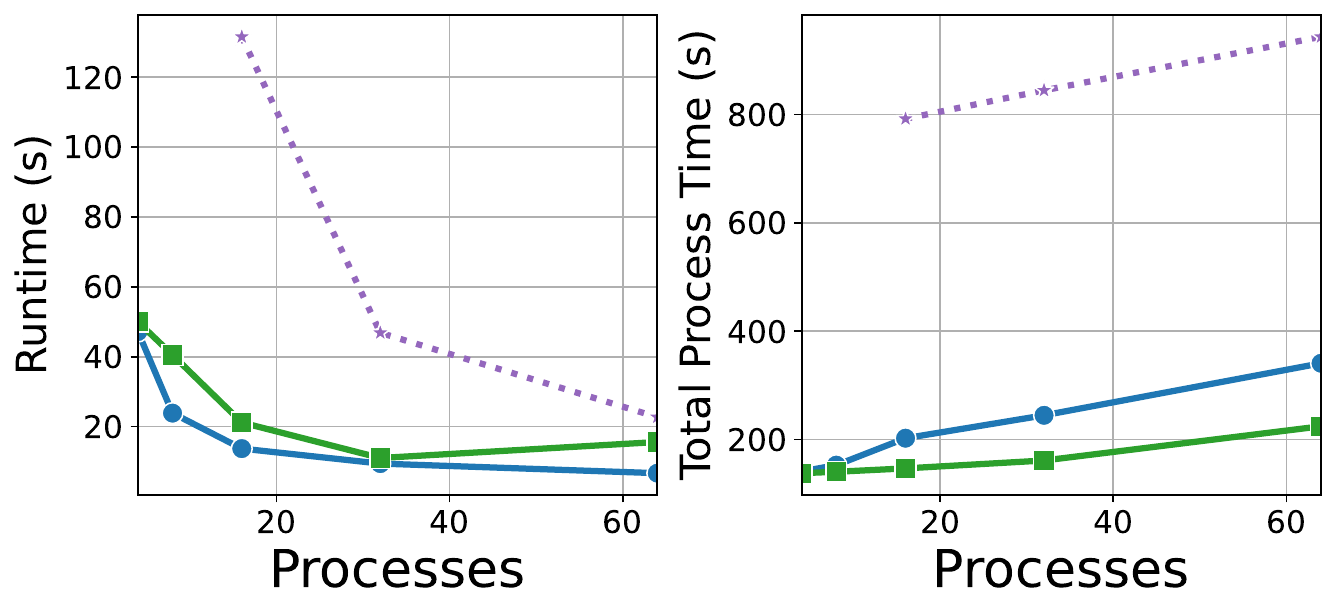}
         \caption{Workload Performance for \wfcorr{} using the \hpc{} with up to 64 processes.}
         \label{fig:corr_hpc}
     \end{subfigure}
    
    \caption{Experiments on \wfcorr{} over different platforms.}
     \label{fig:exp_corr}
\end{figure}

\subsection{Ev. \wfcorr{}}


In the \wfcorr{} workflow, we adopt a consistent workload (50 stations as input) across all platforms. Compared with \wfint{}, \wfcorr{} has more number of PEs and characterises a more heterogeneously distributed workload. It is worth pointing out that, unlike other methods, which start with 4 processes, \multi{} initiates with 12 processes. Since the workflow contains 9 PEs, given the fixed allocation of \multi{}, 9 processes is the minimum requirement. This is a constraint of native static mappings.


As illustrated in Figure~\ref{fig:corr_server} and Figure~\ref{fig:corr_cloud}, \rtime{} of all techniques show a downward trend as the number of processes increases. Conversely, \ptime{} exhibits an increased trend. Notably, the overall performance on \server{} is slightly better than \cloud{} due to the different capacities of these two platforms. The pattern observed includes different mappings and optimizations, various platforms, and different metrics closely aligned with the experiment results from \wfint{}. Given this consistency, we avoid repeatedly mentioning these patterns. Instead, we will conclude with insights from the findings in the subsequent summary section (Section~\ref{subsec:takeaways}).

\subsubsection{Summary of the evaluation on \wfcorr{} with \rtime{} and \ptime{} ratios}


Statistics from Table~\ref{tab:corr} reveal that the overall \ptime{} of this workflow surpasses that of \wfint{}. Notably, while optimal \rtime{} ratios in the previous workflow were generally under 1, they exceed 1 in this case. This indicates that the \auto{} optimization faces challenges with \rtime{} in complex workflows. This could stem from the limitations of the naive auto-scaling algorithm in accurately gauging demand for intricate workflows. However, the \ptime{} ratios, consistently below 1, affirm the efficiency of \auto{} even in complex scenarios.

    
    
    


{
\footnotesize{
\begin{table}[]
\caption{Performance comparison based \wfcorr{} workflow between techniques (A and B). The \rtime{} ratio is the \rtime{} of mapping A over the \rtime{} of mapping B. Similarly, the \ptime{} ratio is the total process time elapsed ratio between A and B. The ratios are prioritized by the metric in the "Prioritized By" column, and the [Mean, Std] shows the average and standard deviation for all \rtime{} and \ptime{} ratios.}
\label{tab:corr}
\begin{tabular}{|l|l|l|l|l|}
\hline
Platform                   & \begin{tabular}{@{}l@{}}Comparison \\ between A/B \end{tabular}        & Prioritized By    & Runtime Ratio     & Process Time Ratio \\ \hline
\multirow{6}{*}{\server{}} & \multirow{3}{*}{$\frac{\automulti{}}{\dynmulti{}}$}  & \rtime{}        & 1.10              & 0.89               \\ \cline{3-5} 
                           &                                                      & \ptime{}        & 1.56              & 0.64               \\ \cline{3-5} 
                           &                                                      & {[}Mean, Std{]} & {[}1.48,  0.31{]} & {[}0.75,  0.11{]}  \\ \cline{2-5} 
                           & \multirow{3}{*}{$\frac{\autoredis{}}{\dynredis{}}$}  & \rtime{}        & 1.25              & 0.98               \\ \cline{3-5} 
                           &                                                      & \ptime{}        & 1.82              & 0.73               \\ \cline{3-5} 
                           &                                                      & {[}Mean, Std{]} & {[}1.61,  0.30{]} & {[}0.91,  0.12{]}  \\ \hline
\multirow{6}{*}{\cloud{}}  & \multirow{3}{*}{$\frac{\automulti{}}{\dynmulti{}}$}  & \rtime{}        & 1.18              & 0.62               \\ \cline{3-5} 
                           &                                                      & \ptime{}        & 1.46              & 0.61               \\ \cline{3-5} 
                           &                                                      & {[}Mean, Std{]} & {[}1.30,  0.13{]} & {[}0.69,  0.14{]}  \\ \cline{2-5} 
                           & \multirow{3}{*}{$\frac{\autoredis{}}{\dynredis{}}$}  & \rtime{}        & 1.05              & 0.90               \\ \cline{3-5} 
                           &                                                      & \ptime{}        & 1.50              & 0.60               \\ \cline{3-5} 
                           &                                                      & {[}Mean, Std{]} & {[}1.47, 0.40{]}   & {[}0.73, 0.13{]}   \\ \hline
\multirow{3}{*}{\hpc{}}    & \multirow{3}{*}{$\frac{\automulti{}}{\dynmulti{}}$}  & \rtime{}        & 1.06              & 0.98               \\ \cline{3-5} 
                           &                                                      & \ptime{}        & 2.34              & 0.66               \\ \cline{3-5} 
                           &                                                      & {[}Mean, Std{]} & {[}1.56, 0.51{]}  & {[}0.79, 0.15{]}   \\ \hline
\end{tabular}
\end{table}
}
}



\subsection{Ev. \wfsent{}}


The goal of this workflow evaluation is to assess \hyredis{}'s performance in stateful applications compared to the baseline \multi{}. For this purpose, the \texttt{happy state} and \texttt{top 3 happiest} stateful PEs have 4 and 2 instances, respectively. With 1 process reserved for stateless PEs, \hyredis{} starts with 8 instances in the experiment. In contrast, \multi{} demands a minimum of 14 processes due to its one-to-one instance-to-process mapping. As the upper process limit is 16, we use finer increments of 8, 10, 12, 14, and 16 processes. 

\begin{figure}
     \centering
     \begin{subfigure}{1\linewidth}
         \centering
         \includegraphics[width=1\linewidth]{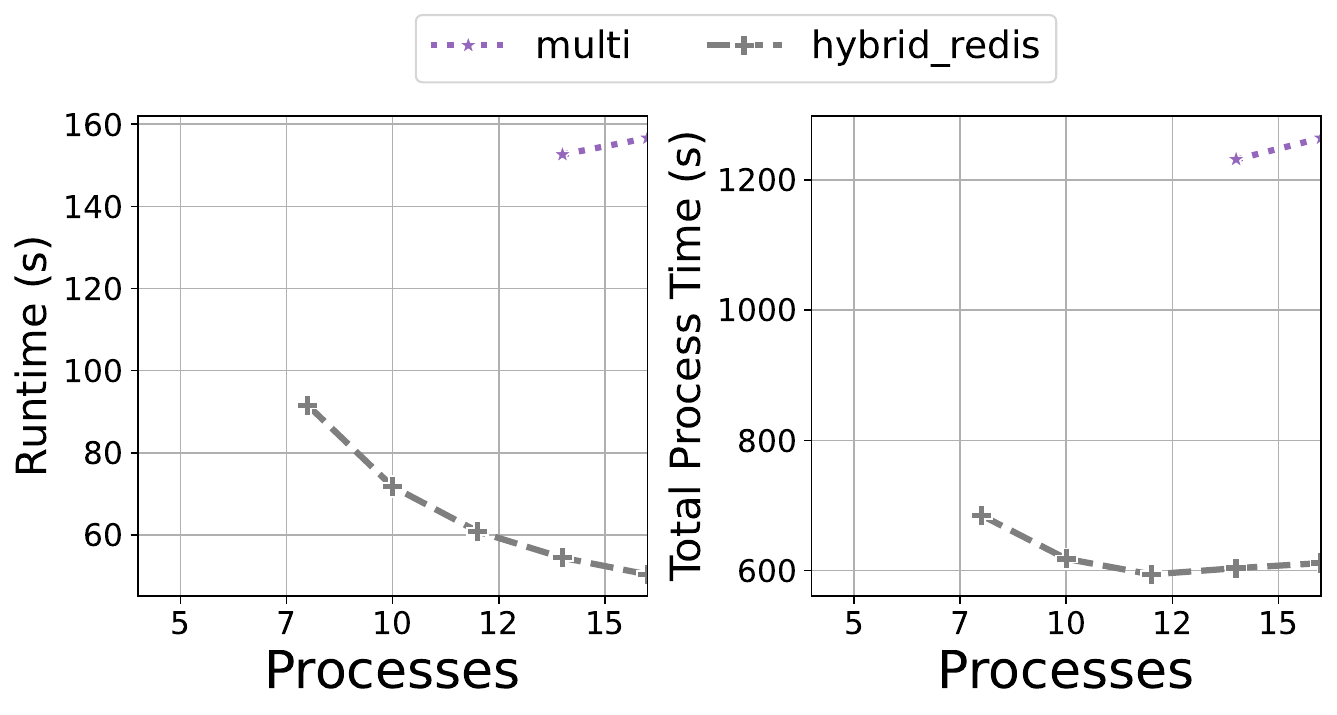}
         \caption{Workload Performance for \wfsent{} using the \server{} with up to 16 processes.}
         \label{fig:sent_int}
    \end{subfigure}
    
    \begin{subfigure}{1\linewidth}
         \centering
         \includegraphics[width=1\linewidth]{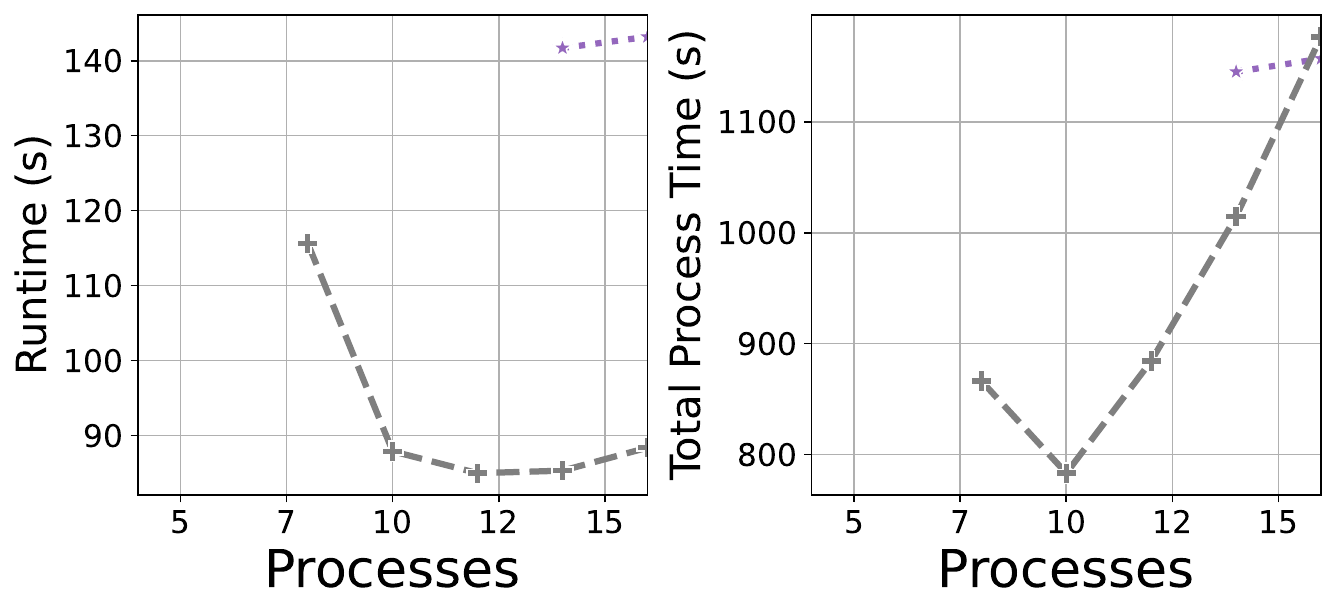}
         \caption{Workload Performance for \wfsent{} using the \cloud{}{} with up to 16 processes.}
         \label{fig:sent_cloud}
    \end{subfigure}
    
    \caption{Experiments on \wfsent{} over different platforms.}
     \label{fig:exp_sent}
     \vspace{-15pt}
\end{figure}

The results on the \server{}, as shown in Figure~\ref{fig:sent_int}, demonstrates that \hyredis{} significantly outperforms the \multi{} in terms of \rtime{}. Additionally, \hyredis{} exhibits a speed-up as the number of processes increases. This can be attributed to the fact that as more processes are added, the stateless workload can be shared by more processes, thereby reducing the overall execution time. Unexpectedly, the trend of \ptime{} is similar to \rtime{}. Given that \auto{} is not applied into \hyredis{}, this efficiency gain likely comes from the more efficient processing of stateless tasks, which reduces the idle time for stateful processing awaiting outputs from stateless PEs. Thus, even with the additional synchronization overhead caused by the increasing number of processes, this reduction in idle time still results in a net efficiency gain.


For the results on \cloud{} which is limited to 8 cores, the drawbacks of over-allocating processing become evident.  While \hyredis{} shows a downtrend in the \rtime{}, its \ptime{} tells a different story. Since available cores have to keep shifting to support an oversized number of processes, the latency introduced by the switching, especially when stateful instances are left idle, leads to a dramatic increase in total idle time. However, despite these, both \rtime{} and \ptime{} for \hyredis{} remain superior to the \multi{}.

\subsubsection{Summary of the evaluation on \wfsent{} with \rtime{} and \ptime{} ratios}


In terms of the \rtime{} and \ptime{} ratios from various platforms and priorities, all ratios are smaller than 1. This indicates that, with \dyn{} optimization, \hyredis{} outperforms \multi{}. This is especially noteworthy, based on the observation that the \redis{} mapping is overall slower than \multiprocessing{} with the same settings.

{
\footnotesize{
\begin{table}[]
\caption{Performance comparison based \wfsent{} workflow between techniques (A and B). The \rtime{} ratio is the \rtime{}  of mapping A over the \rtime{}  of mapping B. Similarly, the \ptime{} ratio is the total process time elapsed ratio between A and B. The ratios are prioritized by the metric in the "Prioritized By" column, and the [Mean, Std] shows the average and standard deviation for all \rtime{} and \ptime{} ratios.}
\label{tab:sent}
\begin{tabular}{|l|l|l|l|l|}
\hline
Platform                   & \begin{tabular}{@{}l@{}}Comparison \\ between A/B \end{tabular}  & Prioritized By    & Runtime Ratio     & Process Time Ratio \\ \hline
\multirow{3}{*}{\server{}} & \multirow{3}{*}{$\frac{hybrid\_redis}{multi}$} & \rtime{}        & 0.32              & 0.48               \\ \cline{3-5} 
                           &                                                & \ptime{}        & 0.32              & 0.48               \\ \cline{3-5} 
                           &                                                & {[}Mean, Std{]} & {[}0.34,  0.02{]} & {[}0.49,  0.01{]}  \\ \hline
\multirow{3}{*}{\cloud{}}  & \multirow{3}{*}{$\frac{hybrid\_redis}{multi}$} & \rtime{}        & 0.60              & 0.89               \\ \cline{3-5} 
                           &                                                & \ptime{}        & 0.60              & 0.89               \\ \cline{3-5} 
                           &                                                & {[}Mean, Std{]} & {[}0.61,  0.01{]} & {[}0.95,  0.09{]}  \\ \hline
\end{tabular}
\end{table}
}
}

\begin{figure*}
     \centering
     \begin{subfigure}[b]{0.3\textwidth}
         \centering
         \includegraphics[width=1\textwidth]{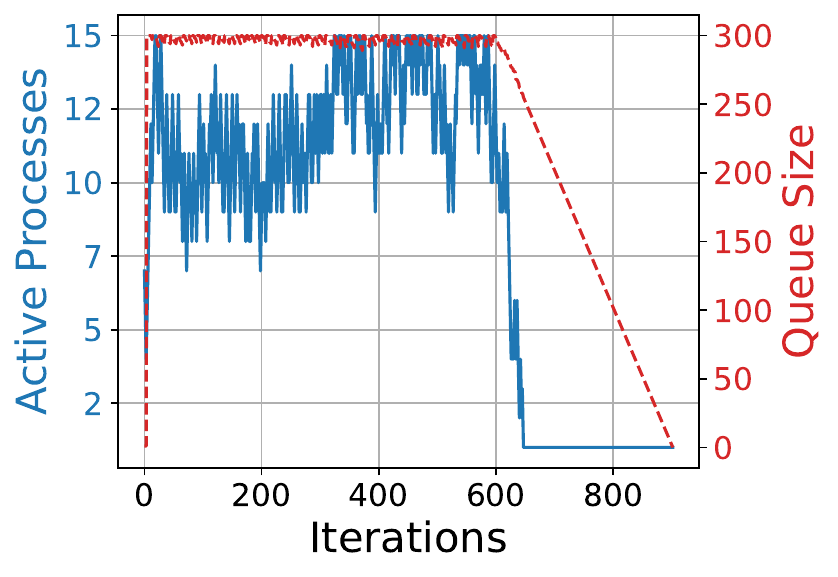}
         \caption{\wfint{} running on \server{} with \automulti{}.}
         \label{fig:exp_auto_scaler_int_server_multi}
     \end{subfigure}
     \begin{subfigure}[b]{0.3\textwidth}
         \centering
         \includegraphics[width=1\textwidth]{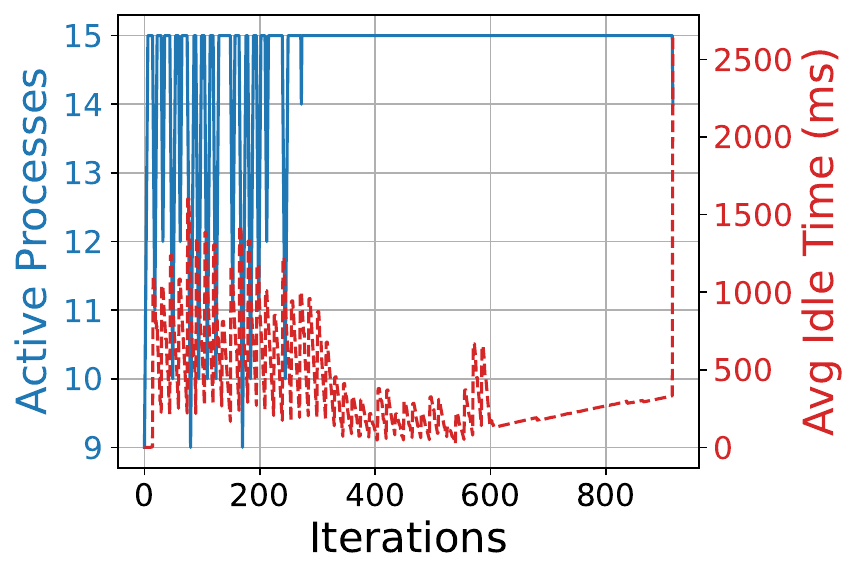}
         \caption{\wfint{} running on \server{} with \autoredis{}.}
         \label{fig:exp_auto_scaler_int_server_redis}
     \end{subfigure}
     \begin{subfigure}[b]{0.3\textwidth}
         \centering
         \includegraphics[width=1\textwidth]{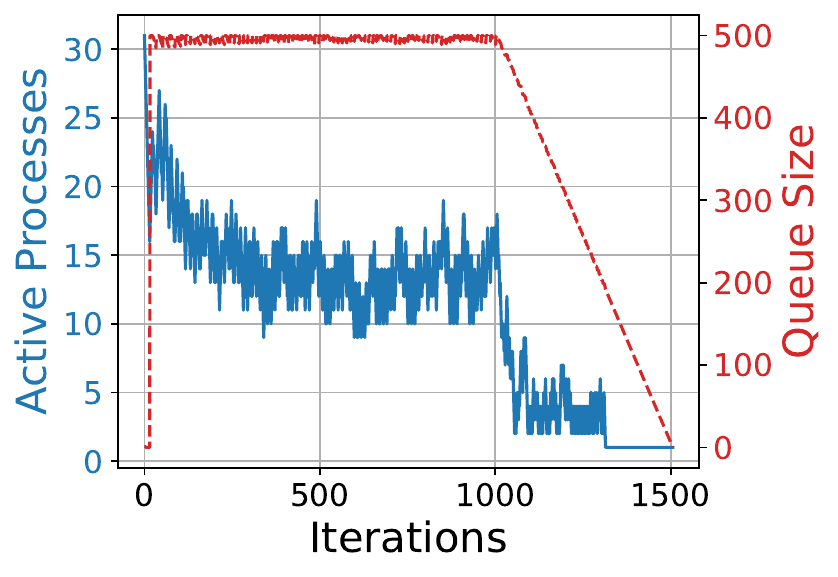}
         \caption{\wfint{} running on \hpc{} with \automulti{}.}
         \label{fig:exp_auto_scaler_int_hpc_multi}
     \end{subfigure}
    \begin{subfigure}[b]{0.3\textwidth}
         \centering
         \includegraphics[width=1\textwidth]{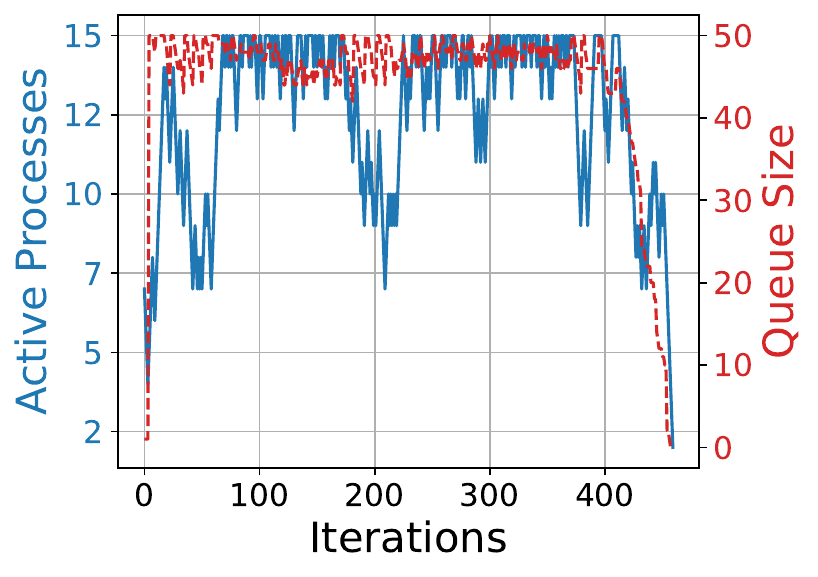}
         \caption{\wfcorr{} running on \server{} with \automulti{}.}
         \label{fig:exp_auto_scaler_corr_server_multi}
     \end{subfigure}
     \begin{subfigure}[b]{0.3\textwidth}
         \centering
         \includegraphics[width=1\textwidth]{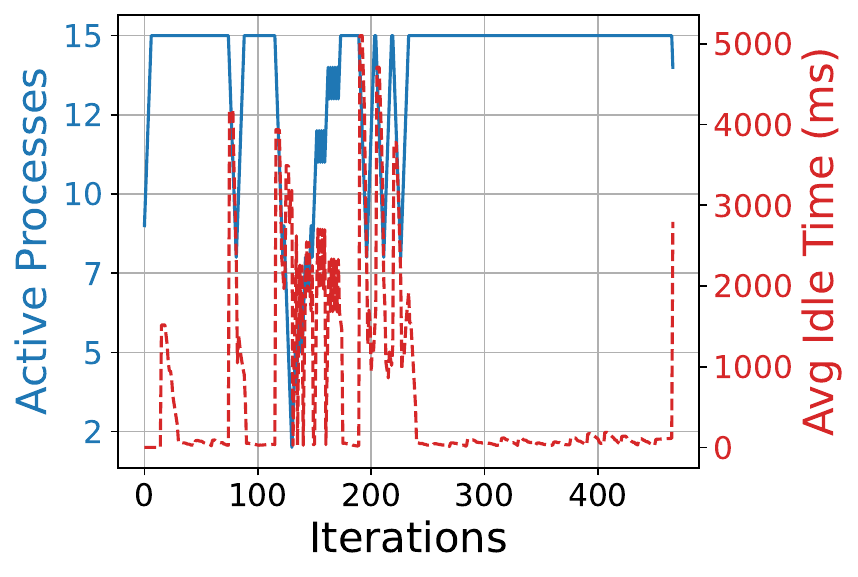}
         \caption{\wfcorr{} running on \server{} with \autoredis{}.}
         \label{fig:exp_auto_scaler_corr_server_redis}
     \end{subfigure}
     \begin{subfigure}[b]{0.3\textwidth}
         \centering
         \includegraphics[width=1\textwidth]{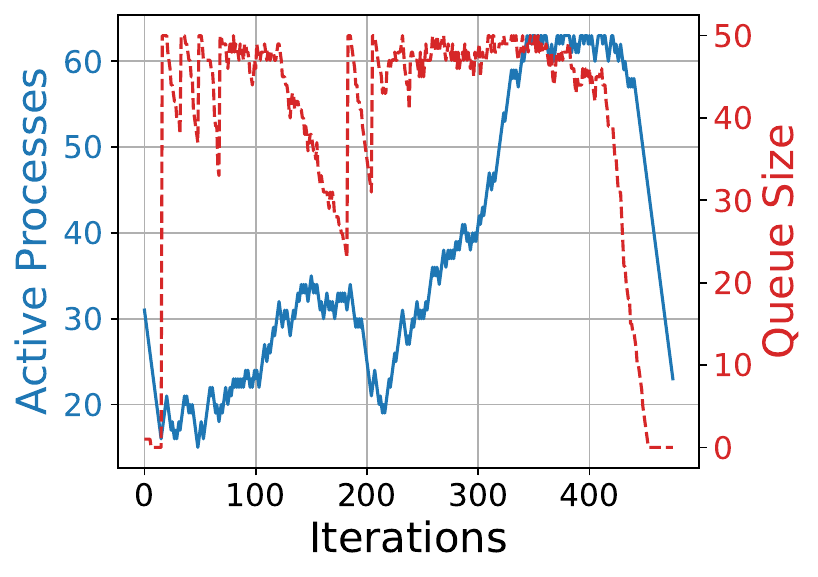}
         \caption{\wfcorr{} running on \hpc{} with \automulti{}.}
         \label{fig:exp_auto_scaler_corr_hpc_multi}
     \end{subfigure}

     \caption{Active Size and the monitoring metrics (Queue Size, and Idle Time) running with  \wfint{} workflow and \wfcorr{} workflow.}
     \label{fig:exp_auto_scaler}
\end{figure*}

\subsection{Analysis on \auto{}}

We experiment with the \textit{auto-scaler} mechanism using the \wfint{} and \wfcorr{} workflows on both \server{} and \hpc{}. In Figure~\ref{fig:exp_auto_scaler}, the left y-axis depicts the active process count throughout runtime. The right y-axis represents the queue size for \automulti{} and the average idle time for \autoredis{}. The x-axis indicates iteration counts, recorded when monitored metrics (right y-axis) change. Please note that these iterations are not uniformly spaced in terms of time interval.

For \automulti{} in Figure~\ref{fig:exp_auto_scaler_int_server_multi}, \ref{fig:exp_auto_scaler_int_hpc_multi}, \ref{fig:exp_auto_scaler_corr_server_multi} and \ref{fig:exp_auto_scaler_corr_hpc_multi}, there is noticeable positive correlation between the number of active processes and queue size. It is in line with our expectations: a larger workload in the queue needs more active processes. Notably, in \hpc{}, especially, for the simple workflow, the active size rarely reaches the maximum limit (64), even though the queue size consistently remains at a high level. This is probably because the naive \textit{auto-scaler} of \automulti{} adjusts the active size only by considering the changes in the queue size without the absolute workload.

For \autoredis{} in Figure~\ref{fig:exp_auto_scaler_int_server_redis} and Figure~\ref{fig:exp_auto_scaler_corr_server_redis}, there is an inverse relationship between the number of active processes and the average idle time. This means that the \textit{auto-scaler} reduces the active size when active processes have larger idle periods, indicating a reduced workload. The sub-figures reveal a consistent trend: active process numbers lag behind metric changes due to inertia in the naive \auto{} strategy. This can result in mismatches between actual needs and active process count. We recognize the need for optimizing the auto-scaling strategy, enhancing its ability to accurately capture real requirements and predict workload changes.

\subsection{Key Insights}
\label{subsec:takeaways}
Summarizing our extensive experiments, key findings include: 
\begin{itemize}
    \item \textbf{Consistent \auto{} Efficiency}: \auto{} consistently demonstrates efficiency across diverse platforms and workflows. It achieves 87\% \rtime{} and 76\% \ptime{} of \dyn{}'s performance in optimal cases

    \item \textbf{Complex Workflow Challenges}: While generally effective, \auto{} faces challenges with complex workflows. Its auto-scaler can struggle to accurately predict needs, causing slight \rtime{} extensions. However, \ptime{} efficiency remains strong.

    \item \textbf{Stateful Mapping Superiority}: In the context of workflows involving stateful applications, \hyredis{} surpasses its counterpart \multi{}, achieving as low as 32\% of \rtime{} compared to \multi{}.

    \item \textbf{\multiprocessing{} vs. \redis{} Performance}: The performance achieved with the \textit{Multiprocessing} optimizations (\dynmulti{} and \automulti{})  outperforms those of \textit{Redis} (\dynredis{} and  \autoredis{}), primarily attributed to the lightweight nature of \multiprocessing{} - despite employing the same optimization strategies for both \textit{Multiprocessing} and \textit{Redis}.





\end{itemize}

\section{Conclusions}
\label{sec:concl}

This paper presents our work harnessing the capabilities of the \redis{} framework to incorporate the \dyn{} optimization, complemented by a new \auto{} strategy. Furthermore, we have expanded the horizon of \dyn{} optimizations to accommodate stateful applications through the introduction of a novel \hyredis{} approach. Through  experiments across diverse workflows and platforms, we demonstrated that \auto{} achieves efficiency while preserving performance in most cases. Notably, the stateful optimization (\hyredis{}) outperformed native mappings. While initial \auto{} strategies have limitations, our study lays a foundation for refining and enhancing auto-scaling within \dfp{}. This integration  benefits \dfp{} and offers insights for enhancing other scientific workflows. Our contributions pave the way for future advancements and provide valuable insights to the scientific community.




\bibliographystyle{ACM-Reference-Format}
\bibliography{reference.bib}

\appendix









\end{document}